\documentclass[12pt,reprint,aps,groupedaddress,nofootinbib,prd,twocolumn]{revtex4-2}

\usepackage[english]{babel}
\usepackage[utf8]{inputenc}
\usepackage{amsmath}
\usepackage{mathbbol}
\usepackage{amssymb}
\usepackage{bbold}
\usepackage{graphicx,amsfonts}
\usepackage{epsfig}
\usepackage{xcolor}
\definecolor{crimsonglory}{rgb}{0.75,0.0,0.2}
\usepackage[colorlinks=true,
linkcolor=crimsonglory,
urlcolor=crimsonglory,
citecolor=crimsonglory]{hyperref}
\usepackage{bm}
\usepackage{mathrsfs}
\usepackage{enumerate}
\usepackage{amsthm}
\usepackage{bbm}
\usepackage{comment}
\usepackage{physics}
\usepackage{url}
\usepackage{upgreek}
\usepackage[title]{appendix}
\usepackage{amssymb,amsmath,amsfonts}
\usepackage{graphicx}
\usepackage{xcolor}

\begin{document}
\date{\today}
\title{Resonant excitation of eccentricity in spherical extreme-mass-ratio inspirals}
\author{Areti Eleni$^{1}$, Kyriakos Destounis$^{2,3,4}$, Theocharis A. Apostolatos$^{5}$ and Kostas D. Kokkotas$^{4}$}
\affiliation{$^1$Department of Mathematics, School of Natural Sciences, University of Patras, 26504 Patras, Greece}
\affiliation{$^2$Dipartimento di Fisica, Sapienza Università di Roma, Piazzale Aldo Moro 5, 00185, Roma, Italy}
\affiliation{$^3$INFN, Sezione di Roma, Piazzale Aldo Moro 2, 00185, Roma, Italy}
\affiliation{$^4$Theoretical Astrophysics, IAAT, University of T\"ubingen, 72076 T\"ubingen, Germany}
\affiliation{$^5$Section of Astrophysics, Astronomy, and Mechanics, Department of Physics, National and Kapodistrian University of Athens, Panepistimiopolis Zografos GR15783, Athens, Greece}

\begin{abstract}
Gravitational radiation reaction, has been one of the fundamental issues in general relativity. Over a span of decades, this process has been analyzed in the adiabatic limit, in order to comprehend how it drives extreme-mass-ratio binaries, that are prime targets for space-borne detectors. It has been shown that spherical orbits around Schwarzschild and Kerr black holes remain spherical (zero eccentricity) under the influence of gravitational radiation reaction. Here, we show that spherical orbits in non-Kerr black holes, that still preserve most of the good qualities and symmetries of Kerr spacetime, can access certain resonances in such a way that an initially spherical inspiral acquires non-zero eccentricity and becomes non-spherical. Therefore, the crossing of resonances under radiation reaction interrupts and even inverts, up to some small radius close to plunge, the process of circularization of orbits. The strength of resonant excitation of eccentricity depends on the initial position and inclination of the integrable extreme-mass-ratio system, as well as the integrability-breaking parameter introduced in the background spacetime that amplifies further the excitation. We find that the harmonics of gravitational waves emitted from these inspirals undergo a frequency modulation as the orbit `metamorphoses' from spherical to non-spherical, due to the effect of resonant eccentricity excitation. The gain that low-amplitude harmonics experience in these oligochromatic EMRIs, due to resonances, may be detectable with future spaceborne detectors and serves as an indicator of non-Kerrness of the background spacetime.
\end{abstract}

\maketitle

\section{Introduction}

Gravitational-wave (GW) observations from the LIGO-Virgo-KAGRA (LVK) collaboration \cite{KAGRA:2021vkt,KAGRA:2023pio} have been paramount in understanding the relativistic aspects of strong-field binary coalescence, and especially the merger of black holes (BHs) and/or neutron stars (NSs). The LVK collaboration is a conglomerate of ground-based, laser interferometers that can detect BH, BH-NS, and NS binaries in the few tens to hundreds of $\text{Hz}$ GW regime. Future, spaceborne, detectors will give access to the $\text{mHz}$ GW regime, where a plethora of event reside, such as supermassive BH binaries, galactic white-dwarf mergers and extreme-mass-ratio inspirals (EMRIs), to name a few.

The Laser Interferometer Space Antenna (LISA) \cite{amaroseoane2017laser}, which was very recently adopted by the European Space Agency, is the primary space-based detector, which will be launched approximately in a decade, and will grant us previously-uncharted GW access to the strong-field $\text{mHz}$ range \cite{Barausse:2020rsu,LISA:2022yao,LISA:2022kgy,Karnesis:2022vdp}, where EMRIs reside. EMRIs consist of a primary, supermassive compact object (usually a BH), and a secondary, stellar-mass compact object, that traces the strong-field regime of the primary due to their vast mass disparity. In turn, since the secondary is $4$ to $7$ orders of magnitude smaller than the supermassive primary, perturbative techniques \cite{Pani:2013pma} can be applied to approximate the inspiral stage of an extreme-mass-ratio system \cite{Babak:2017tow,Berry:2019wgg}.

EMRIs are expected to perform $10^{4}-10^{5}$ revolutions in the strong-field regime, due to their adiabatic evolution, thus will, in principle, trace the background spacetime of the primary and open a new domain of GW phenomenology \cite{Glampedakis:2005hs,Amaro-Seoane:2007osp,Cardenas-Avendano:2024mqp} regarding the validity of general relativity (GR) \cite{Glampedakis:2005cf,Barack:2006pq,Gair:2011ym,Vigeland:2011ji,Cardoso:2011xi,Yunes:2011aa,Pani:2011xj}, the existence of dark compact objects \cite{Macedo:2013jja,Maggio:2021ans,Maggio:2021uge,Cardoso:2022fbq,Maggio:2023fwy}, the effect of dark matter \cite{Macedo:2013qea,Hannuksela:2019vip,Annulli:2020lyc,Kavanagh:2020cfn,Coogan:2021uqv,Figueiredo:2023gas} and astrophysical environments around BHs \cite{Barausse:2014tra,Cardoso:2019rou,Zwick:2022dih,Cardoso:2021wlq,Cardoso:2022whc,Destounis:2022obl,Strateny:2023edo,Polcar:2022bwv,Cole:2022yzw,Feng:2022evy,Stelea:2023yqo}, accretion physics \cite{Babichev:2012sg,Yunes:2011ws,Kocsis:2011dr}, superradiant clouds \cite{Cardoso:2005vk,Witek:2012tr,Brito:2014wla,Brito:2015oca,East:2018glu,Destounis:2019hca,Brito:2020lup,Mascher:2022pku,Destounis:2022rpk,Siemonsen:2022yyf,Brito:2023pyl,Dias:2023ynv}, ultralight fundamental fields \cite{Hannuksela:2018izj,Brito:2017zvb,DeLuca:2021ite,Collodel:2021jwi,Maselli:2020zgv,Maselli:2021men,Barsanti:2022ana,Barsanti:2022vvl}, transient orbital and tidal resonances \cite{Apostolatos:2009vu,Lukes-Gerakopoulos:2010ipp,Contopoulos:2011dz,Flanagan:2010cd,Lukes-Gerakopoulos:2012qpc,Flanagan:2012kg,Ruangsri:2013hra,Berry:2016bit,Zelenka:2017aqn,Bonga:2019ycj,Speri:2021psr,Nasipak:2021qfu,Gupta:2021cno,Gupta:2022fbe,Nasipak:2022xjh,Zelenka:2019nyp,Destounis:2020kss,Destounis:2021mqv,Destounis:2021rko,Lukes-Gerakopoulos:2021ybx,Gutierrez-Ruiz:2018tre,Deich:2022vna,Chen:2022znf,Mukherjee:2022dju,Chen:2023gwm,Destounis:2023gpw,Eleni:2023mjx,Destounis:2023khj,Destounis:2023cim,Lynch:2024ohd} and other relativistic effects \cite{Yunes:2007zp}.

Gravitational self-force (or radiation reaction) \cite{Detweiler:2000gt,Ori:2000zn,Gralla:2008fg,Pound:2010pj,Poisson:2011nh,Barack:2018yvs,Pound:2021qin,Wardell:2021fyy,Burke:2023lno,Lynch:2024ohd} is one of the most successful techniques to evolve EMRIs in order to generate accurate gravitational waveforms \cite{Blanchet:1984wm,Ryan:1995wh,Mino:1996nk,Glampedakis:2002ya,Mino:2003yg,Barack:2002mh,Pound:2012nt,vandeMeent:2017bcc,Upton:2023tcv}. Even so, calculating the self-force of a secondary around a Kerr supermassive primary is an extremely daunting, theory-dependent, and time-consuming task, therefore alternative methodologies have been used to approximate the orbital trajectory and resulting waveform of EMRIs. To that end, a variety of approximations to the process of radiation reaction have been constructed and used over the years, with great success \cite{Burke:1970wx,Shibata:1994jx,Glampedakis:2002cb,Lopez-Aleman:2003sib,Barack:2003fp,Hughes:2005qb,Sopuerta:2005rd,Babak:2006uv,Sundararajan:2007jg,Sundararajan:2008zm,Yunes:2009ef,Yunes:2010zj,Gair:2005ih,Sopuerta:2011te,Chua:2017ujo,Chua:2020stf,Katz:2021yft}.

The influence of gravitational radiation reaction on orbiting particles around BHs is of great interest, especially in EMRIs \cite{Apostolatos:1993nu,Hughes:1999bq,Hughes:2001jr,Pugliese:2010ps,Pugliese:2011py,Pugliese:2011xn,Das:2016opi,Teo:2020sey}. It has been shown that at the adiabatic limit, radiation reaction drives a particle in spherical motion around a Schwarzschild \cite{Apostolatos:1993nu} or a Kerr BH \cite{Ryan:1995xi,Kennefick:1995za} through successively-damped spherical geodesics, i.e. spherical orbits of particles remain spherical (with zero eccentricity) under the action of gravitation radiation reaction up to some minimum radius which is quite close to the separatrix (the analogue of ISCO for non-equatorial orbits). The stability of spherical Kerr orbits has been proven by Kennefick and Ori \cite{Kennefick:1995za} who calculated the relation between the rates of change of energy, of the axial component of angular momentum and that of the Carter constant $Q$ \cite{Carter:1968rr} under a generic gravitational self-force (GSF). They demonstrate that the rate of change of $Q$ has the ``appropriate'' form so that spherical orbits remain spherical at adiabatic order so long as the self-force does not resonate with the radial oscillations. Because the periodicity of the GSF for a spherical orbit is determined by the polar motion and due to the reflection symmetry of the metric, the frequency of GSF is twice the polar frequency. Thus the only assumption for this stability to hold is the fact that there is no resonance of the form $T_\theta=2n T_ r$, for some integer $n$, where $T_\theta$ is the $\theta$-motion period and $T_ r$ is the period of the small-oscillation radial period \cite{Kennefick:1995za,Ryan:1995xi}. This resonance condition is never met in Kerr geodesics. In spite of that, the circularity preservation of EMRIs under radiation reaction has been proven only for vacuum solutions of GR, such as the Kerr family \cite{Kerr:1963ud,Boyer:1966qh}, which have integrable geodesics \cite{Contopoulos:2002}. Nevertheless, the satisfaction of the resonance condition maybe occur in other metrics.

A natural question arises from the above discussion: \emph{Is it possible for the geodesics around  non-Kerr spacetimes to satisfy the resonance condition? If so, will the eccentricity be excited due to the resonant motion?} The aforementioned questions can in principle be addressed by using BHs resulting from modified theories of gravity, BHs surrounded by astrophysical environments or exotic compact objects. Nevertheless, such task is solution-dependent. In this work, we will utilize a parameterized `bumpy' metric, derived by Johannsen \cite{Johannsen}, that describes a dark compact object, for a particular range of deformation parameters, and still possesses a Carter-like constant.\footnote{For the most general case of Kerr-like metrics admitting a Carter constant see \cite{2018CQGra..35r5014P}.} Thus, geodesics are still completely integrable; however the volume of the parameter space where bound geodesic motion occurs can be significantly larger. Consequently, the resonant condition could be satisfied, leading to a resonant excitation of eccentricity when the orbit crosses through a resonance. Since the metric is theory-agnostic, we can in principle tweak the deformation parameters in order to mimic a vast majority of the aforementioned compact objects \cite{Suvorov:2020bvk}. For completeness, we will also consider a modification of Johannsen's metric used in \cite{Destounis:2020kss} that includes an integrability-breaking parameter in order to further examine the `strength' of eccentricity excitation.

We find that, both integrable and non-integrable, non-Kerr EMRIs can satisfy the resonance condition $T_\theta=2n T_r$ for a wide range of deformation parameters. At the orbital level, we observe that a plethora of initial conditions lead to \emph{single} resonant crossings with stronger (or weaker) excitation of eccentricity. The maximization of eccentricity excitation is observed at a particular range of the orbital phase space where the resonance condition is satisfied \emph{twice}, at very close time instants. \emph{Thus, non-Kerr EMRIs possess a resonant-driven mechanism that interrupts, and even inverts for a particular amount of time that depends on the resonance-crossing timescale, the circularization of orbits due to radiation reaction.} The aforementioned phenomenon is imprinted in the GWs of non-Kerr EMRIs as a \emph{resonance-crossing frequency modulation}, i.e. the higher harmonics of the otherwise oligochromatic EMRI gain (at least order of magnitude) amplitude on the expense of the amplitude of the fundamental harmonic. We argue that the potential detection of such type of frequency modulation in EMRIs can point towards a spherical to non-spherical transition when resonances are crossed and thus to a potential `smoking-gun' of non-Kerrness. In what follows, we adopt the geometric units, so that $G=c=1$.

\section{The metric}

We are investigating the evolution of initially spherical orbits assuming that the spacetime geometry of the central dark compact object is described by the metric derived by Johannsen \cite{Johannsen} that includes an extra deformation parameter $a_Q$ \cite{Destounis:2020kss}, which in Boyer-Lindquist-like coordinates $(t,r,\theta,\phi)$ reads:
\begin{equation}
ds^2=g_{tt}dt^2+2g_{t\phi}dt d\phi+g_{rr}dr^2+g_{\theta\theta}d\theta^2+g_{\phi\phi}d\phi^2,
\end{equation}
where
\begin{equation}\label{metric_tensor}
\begin{aligned}
    g_{tt}&=-\frac{\Tilde{\Sigma}[M^3 (a_Q/r)+\Delta-a^2A_2^2(r)\sin^2{\theta}]}{[(r^2+a^2)A_1(r)-a^2A_2(r)\sin^2{\theta}]^2},\\
    g_{t\phi}&=-\frac{a[(r^2+a^2)A_1(r)A_2(r)-\Delta]\Tilde{\Sigma}\sin^2{\theta}}{[(r^2+a^2)A_1(r)-a^2A_2(r)\sin^2{\theta}]^2},\\
    g_{rr}&=\frac{M^3 (a_Q/r)+\Tilde{\Sigma}}{\Delta A_5(r)},\\
    g_{\theta\theta}&=\Tilde{\Sigma},\\
    g_{\phi\phi}&=\frac{\Tilde{\Sigma}\sin^2{\theta}[(r^2+a^2)^2A_1^2(r)-a^2\Delta\sin^2{\theta}]}{[(r^2+a^2)A_1(r)-a^2A_2(r)\sin^2{\theta}]^2},
\end{aligned}
\end{equation}
and 
\begin{equation}
    \Delta=r^2+a^2-2M r,\qquad \Tilde{\Sigma}=r^2+a^2\cos^2{\theta}+f(r).
\end{equation}
The parameters $a$ and $M$, are the spin and the mass of central object, respectively. We assume that the deviation functions are given by the following expressions:
\begin{align}
    A_1(r)&=1+a_{13}\left(\frac{M}{r}\right)^3,\\
    A_2(r)&=1+a_{22}\left(\frac{M}{r}\right)^2,\\
    A_5(r)&=1+a_{52}\left(\frac{M}{r}\right)^2,\\
    f(r)&=\epsilon_{3}\frac{M^3}{r}.
\end{align}
For the integrable metric, i.e. $a_Q=0$, the deviation parameters must satisfy the following conditions

\begin{equation*}
a_{13}\, , \epsilon_3 >-\kappa^3 \quad \mbox{and} \quad a_{22}\, , a_{52} > -\kappa^2
    \end{equation*}
where
    \begin{equation*}
        \kappa =1+\sqrt{1-\frac{a^2}{M^2}}
    \end{equation*}
in order for the metric to be regular outside of the event horizon, i.e. no violation of the Lorentzian signature and absence of closed timelike curves. Finally, the deformation parameter $a_Q$, first introduced in \cite{Destounis:2020kss}, explicitly breaks the integrability of geodesics and leads to chaotic phenomena around resonances. When all deviation parameters are set to zero Eq. \eqref{metric_tensor} reduces to the Kerr metric.

\section{Fundamental frequencies}

The Johannsen metric, for which $a_Q=0$, has been constructed, as a deviation from Kerr, in order to maintain the separability of the Hamilton-Jacobi equations; that is the integrability of geodesic motion of particles. Consequently, it possesses except for the time-independent Hamiltonian $H=\frac{1}{2}g^{\alpha \beta}p_{\alpha}p_{\beta}$, the energy $E$ and the azimuthal angular momentum $L_z$ (due to stationarity and axisymmetry of the metric, respectively), an additional  first integral of motion, i.e. a Carter-like constant $\tilde{Q}$ \cite{Johannsen}:
\begin{equation}
    \tilde{Q}=p_{\theta}^2+a^2\cos^2{\theta}(\mu^2-E^2)+\cot^2{\theta} L_z^2,
\end{equation}
where $\mu$ is the rest mass of the orbiting particle. The first integrals of motion $P_{\alpha}=(H, E, L_z, \tilde{Q})$ are linearly independent, and in involution, since their Poisson brackets satisfy:
\begin{align*}
    \{P_{\alpha},H\}&=0\\
    \{P_{\alpha},P_{\beta}\}&=0
\end{align*}
and the $4-$form $dH\wedge dE \wedge dL_z \wedge d\tilde{Q}$ is non vanishing for generic (eccentric and inclined) bound orbits.

The momenta $p_{\alpha}$ conjugate to Boyer-Lindquist-like coordinates $(t,r,\theta,\phi)$ are defined as \cite{Johannsen}:
\begin{equation}\label{p}
\begin{aligned}
  p_t & =-E,\qquad\qquad\quad\,\,\,\,
  p_{\phi}=L_z,\\
  p_r &=\pm\frac{\sqrt{R(r)}}{\sqrt{A_5(r)}\Delta},\qquad
  p_{\theta} =\pm\sqrt{\Theta(\theta)},
\end{aligned}
\end{equation}
where
\begin{align}
    R(r)&=P(r)^2-\Delta\left(\mu^2(r^2+f(r))+(L_z-aE)^2+\tilde{Q}\right),\label{R}\\
    \Theta(\theta)&=\tilde{Q}-\cos^2{\theta}\left(a^2(\mu^2-E^2)+\frac{L_z^2}{\sin^2{\theta}}\right),\label{U}\\
    P(r)&=E(r^2+a^2)A_1(r)-aL_zA_2(r).    
\end{align}

The geodesic motion is bound in the $r$-domain $r_2\leq r\leq r_1$, with apoapsis radius $r=r_1$ and periapsis radius $r=r_2$ being the two radial turning points, i.e. the two real largest roots of the radial potential $R(r_{1,2})=0$, and in the $\theta$-domain $\theta_{\min}\leq \theta\leq \pi-\theta_{min}$, with $\theta_{min}$ the minimum polar turning point, i.e. $\Theta(\theta_{min})=0$. The radial and polar motion are of libration type, while the azimuthal motion is periodic as a rotation. In 3-dimensional space (after projecting out the time direction) motion lies on an 3-dimensional torus, $T^3$, \cite{1978mmcm.book.....A} which means that it is of compact support. Even though the spatial coordinates are not really periodic, their oscillation (or rotation with respect to $\phi$) correspond to characteristic frequencies of geodesic bound motion, the so-called \emph{fundamental frequencies}. However, the orbit is not bounded in the time direction. 

Since the system is completely integrable, we will apply a generalization of Arnold-Liouville theorem for non-compact invariant manifolds of completely integrable Hamiltonian systems \cite{2003JPhA...36L.101F, 2008PhRvD..78f4028H}, to include the non-compact time-like coordinate in derivation of  the fundamental frequencies. Consequently, the phase space trajectory of the four degrees of freedom dynamical system is diffeomorphic to the product $T^3\cross R$, where $R$ is the set of real numbers. So, we can define a new set of symplectic coordinates, the generalized action-angle variables $(q_\mu,J_\mu)$, where the angle variables $q_{\mu}$ are linear functions of time, and the action variable $J_{\mu}$ corresponds to a fixed vector. Moreover, the angle variables $q_i$, with $i=r,\theta,\phi$, are periodic functions of time with the frequencies related to their periodicities being the fundamental frequencies that one obtains by Fourier analyzing the oscillating (or rotating) time dependence of the coordinates $r,\theta$ and  $\phi$ of a bound geodesic orbit and are relevant to the computation of gravitational radiation emitted by EMRIs assuming adiabatic approximation. 
 
The generating function of the (type-2) canonical transformation $(x^{\mu},p_{\mu})\to(q_{\mu},J_{\mu})$ is the Hamilton's characteristic function $W(x^{\mu},P_{\alpha})$, which is a solution of the Hamilton-Jacobi equation:
\begin{equation}
    \frac{\partial S}{\partial \tau}+H \left(x^{\mu},\frac{\partial S}{\partial x^{\mu}}\right)=0,
\end{equation}
where
\begin{equation}
S(x^{\mu},P_{\alpha},\tau)=-H \tau+ W(x^{\mu},P_{\alpha}),
\end{equation}
with $\tau$ the proper time, $x^{\mu}$ the Boyer-Lindquist-like coordinates $(t, r,\theta,\phi)$, $p_{\mu}$ their conjugate momenta (\ref{p}) and $P_{\alpha}=(H,E,L_z,\tilde{Q})$ the constants of motion. Due to the separability of the Hamilton-Jacobi equation the characteristic function takes the form
\begin{equation}
    W(x^{\mu},P_{\alpha})=-E t+L_z \phi\pm W_r(r) \pm W_{\theta}(\theta),
\end{equation}
where
\begin{eqnarray}
W_r(r) &=& \int^r \frac{\sqrt{R(r)}}{\Delta\sqrt{A_5(r)}}\;dr, \\
W_\theta(\theta) &=& \int^{\theta} \sqrt{\Theta(\theta)} \;d\theta.
\end{eqnarray}
The action variables are defined as (see \cite{1978mmcm.book.....A})
\begin{equation}
J_\mu=\frac{1}{2\pi}\oint p_{\mu}d x^{\mu},\label{J}
\end{equation}
where the integration for $J_i$ is to be carried over a period of oscillation or rotation of $x^i$ \cite{2002CQGra..19.2743S}, while $J_t$ is integrated over a curve of length $2\pi$ \cite{2008PhRvD..78f4028H}. In our case by substituting Eqs. (\ref{p}), the action variables become:
\begin{align}
J_r&=\frac{1}{2\pi}\oint \frac{\sqrt{R(r)}}{\Delta\sqrt{A_5(r)}} dr\label{Jr},\\
J_{\theta}&=\frac{1}{2\pi}\oint\sqrt{\Theta(\theta)}d\theta \label{Jth},\\
J_{\phi}&=\frac{1}{2\pi}\oint p_{\phi}d\phi=L_z\label{Jph},\\
J_t&=\frac{1}{2\pi}\int_0^{2\pi}p_tdt=-E\label{Jt}.
\end{align}
The action variables $J_{\mu}$ are constants of motion, as they depend only on the first integrals $J_{\mu}=J_{\mu}(P_{\alpha})$. By inverting them we can express the integrals $P_{\alpha}$ as functions  of the action variables. In particular the Hamiltonian, which becomes cyclic with respect to angles $q_{\mu}$, can be expressed as $H=P_0=H(J_{\mu})$. The generating function also can be expressed in terms of the Boyer-Lindquist-like coordinates and action variables as
\begin{equation}
    W=W(x^{\mu},J_{\lambda}).
\end{equation}
The transformation equations become
\begin{eqnarray}
    p_{\mu}&=&\frac{\partial{W}}{\partial{x^{\mu}}}(x^{\nu},J_{\lambda}), \\
    q_{\mu}&=&\frac{\partial{W}}{\partial{J_{\mu}}}(x^{\nu},J_{\lambda}),
\end{eqnarray}
while the equations of motion in action-angle variables read
\begin{eqnarray}
\dot{q}_{\nu}&=&\frac{\partial{H}}{\partial{J}_{\nu}}=\Omega_{\nu}(J_{\mu}),\\
\dot{J}_{\nu}&=&-\frac{\partial{H}}{\partial{q}_{\nu}}=0.
\end{eqnarray}
The $(q_r, q_{\theta}, q_{\phi})$ angle variables are periodic and linear functions of time, i.e.
\begin{equation}
q_i(\tau) =\left(\Omega_{i}(J_{\mu}) \tau+ q_i(0)\right)\mod{2 \pi},
\end{equation}
where $\Omega_{i}(J_{\mu})$ and $q_i(0)$ are constants and  $\Omega_{i}(J_{\mu})=\partial{H}/\partial{J}_{i}$ describe the fundamental frequencies of the orbit, for $i=r,\theta,\phi$. 

In order to derive the expressions of the frequencies, we should express the Hamiltonian with respect to the action variables $H(J_{\mu})$; a task that can only be performed numerically. The integrals (\ref{Jr})-(\ref{Jt}) of action variables cannot be explicitly inverted. However, we can calculate the frequencies from the inverse derivatives $\partial{J_{\mu}}/\partial{P_{\beta}}$, combined with the chain rule: 
\begin{equation}\label{chain}
    \frac{\partial{P_{\alpha}}}{\partial{J_{\mu}}}\frac{\partial{J_{\mu}}}{\partial{P_{\beta}}}=\delta_{\beta}^{\alpha}.
\end{equation}

The non-trivial partial derivatives read
\begin{align}
\frac{\partial{J}_r}{\partial{H}}&=\frac{Y}{\pi},\label{1}\\
\frac{\partial{J}_r}{\partial{E}}&=\frac{W}{\pi},\label{2}\\
\frac{\partial{J}_r}{\partial{L_z}}&=-\frac{Z}{\pi},\label{3}\\
\frac{\partial{J}_r}{\partial{Q}}&=-\frac{X}{2\pi},\label{4}\\
\frac{\partial{J}_{\theta}}{\partial{H}}&=\frac{2\sqrt{z_{+}}a^2}{\pi \beta}[K(k)-E(k)],\label{5}\\
\frac{\partial{J}_{\theta}}{\partial{E}}&=\frac{2\sqrt{z_{+}}E a^2}{\pi \beta}[K(k)-E(k)],\label{6}\\
\frac{\partial{J}_{\theta}}{\partial{L_z}}&=\frac{2L_z}{\pi \beta\sqrt{z_{+}}}[K(k)-\Pi(z_{-},k)],\label{7}\\
\frac{\partial{J}_{\theta}}{\partial{Q}}&=\frac{1}{\pi \beta\sqrt{z_{+}}}K(k).\label{8}
\end{align}
The  $Y$, $W$, $Z$ and $X$ are the radial integrals
\begin{align}
Y&=\int^{r_1}_{r_2}\frac{r^2+f(r)}{\sqrt{A_5(r)R(r)}}dr,\label{Y}\\
W&=\int^{r_1}_{r_2}\frac{dr}{\Delta \sqrt{A_5(r)R(r)}}\left[(r^2+a^2)A_1(r)\times\right.\nonumber\\
&\left.[(r^2+a^2)A_1(r)E-a A_2(r)L_z]+\Delta a(L_z-a E)\right],\label{W}\\
Z&=\int^{r_1}_{r_2}\frac{dr}{\Delta\sqrt{A_5(r)R(r)}}\left[a 
 A_2(r)\times\right.\nonumber\\
&\left.[(r^2+a^2)A_1(r)E-a A_2(r)L_z]+\Delta(L_z-a E)\right],\label{Z}\\
X&=\int^{r_1}_{r_2}\frac{dr}{\sqrt{A_5(r)R(r)}},\label{X}
\end{align}
where the apoapsis $r_1$ and periapsis $r_2$ are the turning points of the radial motion, i.e. $R(r_1)=R(r_2)=0$.

The $K(k)$, $E(k)$ and $\Pi(z_{-},k)$ are the 1st, 2nd and 3rd complete elliptic integrals, respectively \cite{1972hmfw.book.....A}:
\begin{align}\label{ElK}
K(k)&=\int^{\frac{\pi}{2}}_0
\frac{d\theta}{\sqrt{1-k^2\sin^2{\theta}}},\\
\label{ElE}
E(k)&=\int^{\frac{\pi}{2}}_0
\sqrt{1-k^2\sin^2{\theta}}\;d\theta,\\
\label{ElP}
\Pi(z_{-},k)&=
\int^{\frac{\pi}{2}}_0
\frac{d{\theta}}{(1-z_{-}\sin^2{\theta})
\sqrt{1-k^2\sin^2{\theta}}},
\end{align}
with $z=\cos^2{\theta}$, $k=\sqrt{z_{-}/z_{+}}$ (where $z_{\pm}$ are the two roots of $\Theta(z)=0$ with $0<z_-=\cos^2{\theta_{min}}<1<z_+$) and $\beta^2=a^2(\mu^2-E^2)$.

Substituting the derivatives (\ref{1})-(\ref{8}) to the chain rule (\ref{chain}) and solving with respect to $\partial H/\partial J_{\mu}$, we obtain the desired fundamental frequencies, which are given by the following expressions:
\begin{align}
\Omega_t&=\frac{K(k)W+a^2z_+E[K(k)-E(k)]X}{a^2z_+[K(k)-E(k)]X+K(k)Y},\label{wt}\\
\Omega_r&=\frac{\pi K(k)}{a^2z_{+}[K(k)-E(k)]X+YK(k)},\label{wr}\\
\Omega_{\theta}&=\frac{\pi\beta\sqrt{z_{+}}X/2}{a^2z_{+}[K(k)-E(k)]X+YK(k)},\label{wu}\\
\Omega_{\phi}&=\frac{ZK(k)+XL_z[\Pi(z_{-},k)-K(k)]}{a^2z_{+}[K(k)-E(k)]X+YK(k)}.\label{wf}
\end{align}

The constant $\Omega_t=\frac{\partial H}{\partial J_t}$, associated with the generalized timelike coordinate, cannot be interpreted as a physical fundamental frequency because the motion is not bounded in the timelike direction \cite{2002CQGra..19.2743S}.

Even though the denominators of the integrands in (\ref{Y})-(\ref{X}) vanish at the turning points $r_1$ and $r_2$, the $X$, $Y$, $Z$ and $W$ quantities can be transformed into well-behaved integrals by virtue of the substitution $r=\frac{p}{1+e\cos{\chi}}$, where $p$ is the semi-latus rectum, $e$ is the eccentricity and the variable $\chi$ increases monotonically with time. The turning points then become $r_1=\frac{p}{1-e}$ and $r_2=\frac{p}{1+e}$. We give the corresponding expressions for the radial integrals in $\chi$-representation in Appendix \ref{App.B}.

From (\ref{wr}) and (\ref{wu}) it's obvious that the resonance condition, $\Omega_r/\Omega_{\theta}=2$, is satisfied when
\begin{equation}\label{res. cont.}
    \beta \sqrt{z_+} X=K(k).
\end{equation}
For spherical orbits, i.e. $e=0$,  the turning points become $r_1=r_2= p=r_0$ and the integral $X$ simplifies to the exact expression
\begin{equation}
    X=\frac{\pi}{r_0 \sqrt{(1+a_{52}M^2/r_0^2)J_{\text{circ}}}},
\end{equation}
where the function $J_{\text{circ}}=J_e(\chi)|_{e=0}$ (see Appendix \ref{App.B}) is defined by:
\begin{align*}
    J_{\text{circ}}=& 45 (1-E^2)-72\frac{M}{r_0}+28\frac{a^2(1-E^2)+L_z^2+\
    \tilde{Q}}{r_0^2}\\
    &-42\frac{M}{r_0^3}\left[(L_z-a E)^2+\tilde{Q}-\epsilon_3 M^2/2+a_{13}M^2 E^2\right]\\
    &+15\frac{a^2\tilde{Q}-2M^4\epsilon_3+2a\; a_{22} M^2 E L_z}{r_0^4}\\
    &+10\frac{2a\; a_{13}E M^3(L_z-2 a E)+\epsilon_3 a^2 M^3}{r_0^5}\\
    &-6\frac{a_{13}^2 E^2 M^6+2a_{22}a^2L_zM^2(L_z-a E)}{r_0^6}\\
    &+6\frac{a_{13}a^3 E M^3(L_z-a E)+a\; a_{13}\; a_{22}E L_z M^3}{r_0^7}\\
    &-\frac{2 a^2 a_{13}^2 E^2 M^6+a^2 a_{22}^2 L_z^2 M^4}{r_0^8}.
\end{align*}
The constants of motion $(E,L_z,\tilde{Q})$ should be expressed as functions of the orbital parameters $(e,p,\theta_{\min})$, see Appendix \ref{App. A}.

\section{Initial conditions and inspiral fluxes}

\begin{figure}
    \includegraphics[scale=0.75]{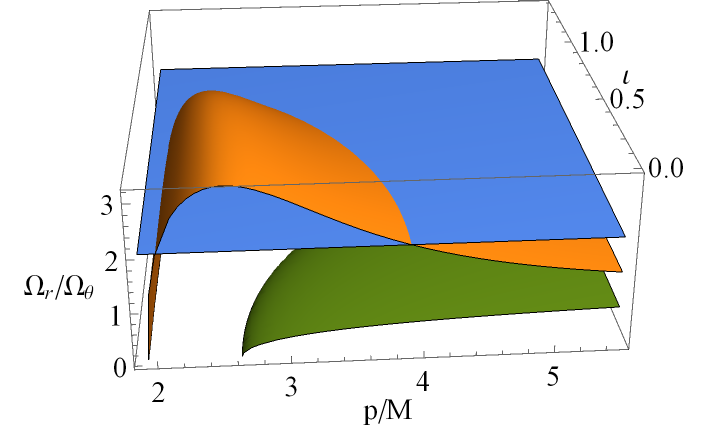}
    \caption{This three-dimensional plot depicts the ratio between the two fundamental frequencies, $\Omega_r/\Omega_\theta$, as a function of the orbital characteristics $p/M$ and $\iota$ of Kerr and non-Kerr Johannsen spherical ($e=0$) orbits with $a/M=0.85$. The non-Kerr metric is characterized by $a_{12}=-1.5$, $a_{22}=4$, $a_{52}=6$ and $\epsilon_3=4$. The orange surface refers to the Johannsen case, while the green one to Kerr. The section between the blue level, marking the condition $\Omega_r/\Omega_\theta=2$, and the orange surface refers to the resonance condition. Therefore while an EMRI spherical orbit evolves towards lower values of $p$ and slightly higher values of $\iota$ along the orange surface, it might hit a resonance twice before it ends up plunging at the separatrix (the leftmost part of the orange surface). The Kerr case (green surface) never hits a resonance.}
    \label{3d-frequency ratio}
\end{figure}

In order to find the initial orbital parameters $(e,\iota,p)$, where $\iota$ is the inclination angle $\iota=\frac{\pi}{2}-\theta_{\min}$ (prograde orbits), for spherical orbits that lie close to the resonance $\Omega_r/\Omega_{\theta}=2$, we set the eccentricity to zero ($e=0$) and choose an inclination angle in the interval $\iota \in [0,\iota_{max}]$, with $\imath_{\max}$ being the angle of the most inclined spherical orbit for which the resonance condition is met. We numerically solve Eq. \eqref{res. cont.} and determine $p$ so that the orbit is at resonance. We note that Eq. \eqref{res. cont.} has two solutions for given $e$ and $\iota$, as shown with the intersection of the blue and orange surfaces in Fig. \ref{3d-frequency ratio}, and even a degenerate solution for large inclination angle and sufficiently small semi-latus rectum. Figure \ref{3d-frequency ratio} also includes the possible solutions of $p/M$ and $\iota$ for which $e=0$ in a Kerr EMRI. It is obvious that all those solutions cannot satisfy the resonance condition, since the green surface never intersects the blue one. We therefore commence by evolving trajectories, with initial conditions slightly close (but larger than) the largest semi-latus rectum that gives a resonant geodesic orbit, by fixing $e=0$ and $\iota \in [0,\iota_{max}]$, while we slowly increase the $p$ till we get a range of resonant-crossing inspirals. For an inclination angle close to $\iota_{\max}$, we start the inspiral with initial inclination slightly lower than the value we used to solve the Eq. \eqref{res. cont.}, otherwise we may miss the resonance as the inclination increases during the inspiral.

To evolve the inspirals we use the fluxes found in \cite{Glampedakis:2002ya,Gair:2005ih} at second post-Newtonian order, i.e. we use the \emph{numerical kludge scheme}, for Kerr and suitably adjust them in order to take into account the deformed quadrupolar structure of spacetime. Here, we note that the Johannsen metric is potentially a non-vacuum solution (see e.g. Ref. \cite{Suvorov:2020bvk}) that explains the difficulty of calculating its multipoles consistently. We have tried a large variety of  different ways of introducing the multipole-related parameters in flux rates and did not find qualitatively different results. For the integrability-breaking case, where $a_Q\neq 0$, we can consistently adjust the fluxes by deforming the mass quadrupole moment $M_2$ of the primary as $M_2=-M a^2-M^3 a_Q/3$, since $a_Q$ appears in the second-order expansion of both $g_{tt}$ and $g_{rr}$ \cite{Destounis:2021mqv}. For all other integrable EMRI evolutions we simply utilize the well-studied Kerr fluxes \cite{Babak:2006uv}\footnote{We note that the leading-order PN fluxes of the constants of motion in integrable EMRIs described by the Johanssen metric ($a_Q=0$ in Eq. \eqref{metric_tensor}) have been found in Ref. \cite{AbhishekChowdhuri:2023gvu}, but only for eccentric equatorial trajectories. The fluxes for inclined eccentric orbit are still unknown.}. 

In what follows, we will evolve integrable and non-integrable EMRIs with the technique discussed above that has been tested thoroughly in \cite{Gair:2007kr,Apostolatos:2009vu,Lukes-Gerakopoulos:2010ipp} and adapted in \cite{Destounis:2021mqv,Destounis:2021rko,Destounis:2023cim,Destounis:2023khj} to include non-linear, updated fluxes every $N$ revolutions around the compact primary. More precisely, we numerically integrate the 2nd order coupled ordinary-differential equations (coupled geodesics) for a small interval of time compared to the radiation reaction timescale, to obtain $r(t)$ and $\theta(t)$, assuming a linear evolution of $E(t)$ and $L_z(t)$. Then we run a geodesic evolution with initial conditions taken from the final point in phase space of the previous adiabatically evolved orbit. From this geodesic orbit we determined numerically the orbital elements $(p, e ,\iota)$ as
\begin{equation}
    e=\frac{r_1-r_2}{r_1 + r_2}, \,\,\,\,\,\,\, p=\frac{2r_1 r_2}{r_1 + r_2},\,\,\,\, \,\,\,\iota=\frac{\theta_1-\theta_2}{2},
\end{equation}
are the maximum and minimum angular values along the geodesic orbit. Finally, we use these orbital elements to update the average losses of $E$ and $L_z$ and we evolve again the orbit under radiation reaction for the next time interval. The time interval of each segment of adiabatic evolution has been adjusted to include $N=10^2-10^3$ orbital revolutions in order to achieve smooth data for the evolution of the orbital elements and eventually the inspiral (see Ref. \cite{Destounis:2021rko} for a detailed presentation on the inspiral evolution scheme).

\section{Eccentricity excitation in integrable EMRIs}

In this section, we treat orbits of integrable EMRIs, that have a Carter-like constant \cite{Johannsen}, with respect to that of Kerr, thus still preserving the integrability properties of the dynamical system and any direct or indirect chaotic behavior is absent. Nevertheless, the spacetime metric of the primary can be sufficiently deformed in order to achieve resonant-crossing spherical orbits. In what follows, we demonstrate that when an initially spherical inspiral, under radiation reaction, crosses the resonance $\Omega_r/\Omega_\theta=2$, then the trajectory is excited into an eccentric one.

\begin{figure*}
    \includegraphics[scale=0.33]{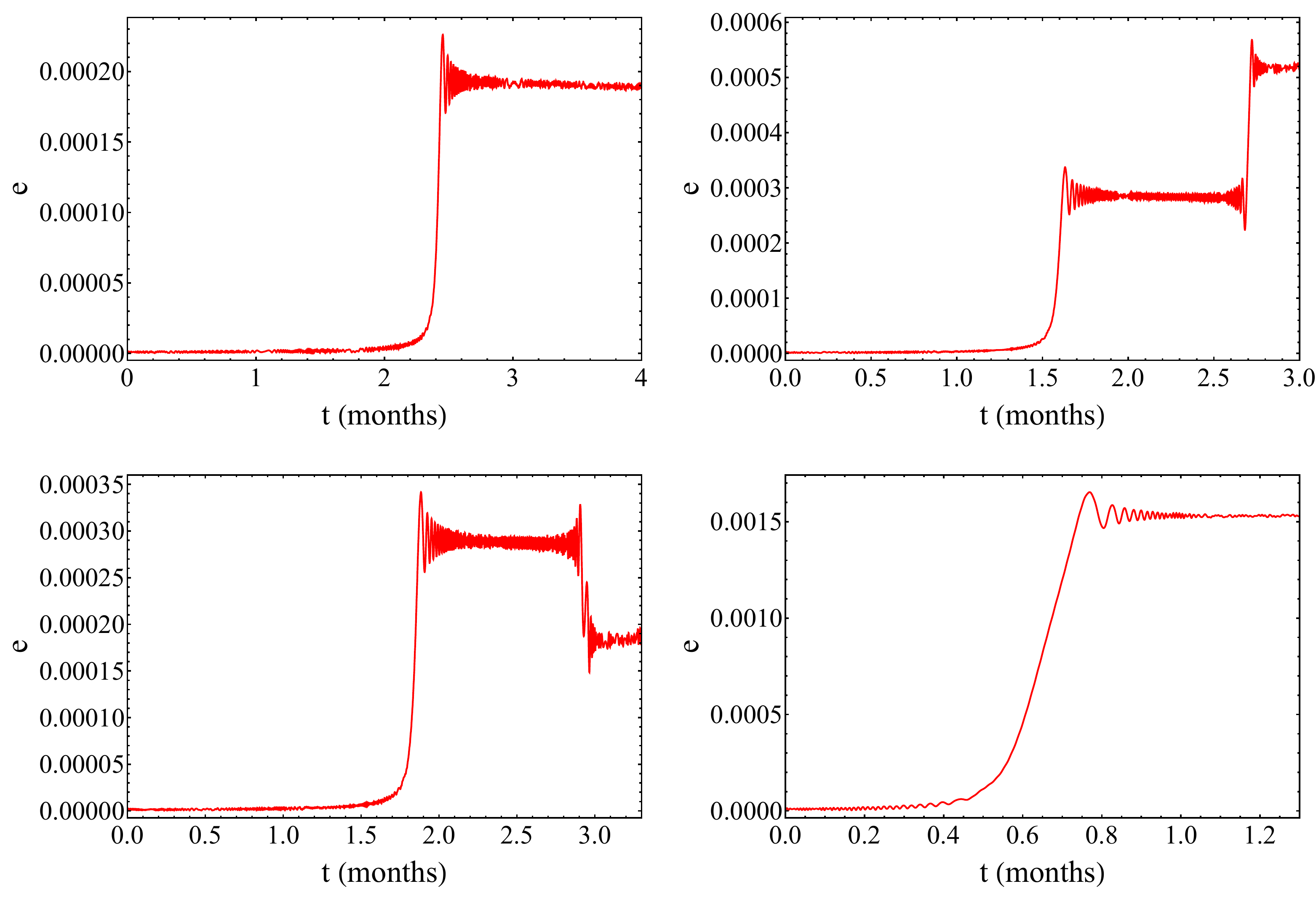}
    \caption{Evolution of a spherical EMRI, where the supermassive primary is described by Eq.~\eqref{metric_tensor}, through the $\Omega_r/\Omega_\theta=2$ resonance. In all cases, we have $\mu/M=10^{-6}$, $a=0.85 M$, $a_{12}=-1.5$, $a_{22}=4$, $a_{52}=6$ and $\epsilon_3=4$. All orbits begin from the equatorial plane, i.e. $\theta(0)=\pi/2$, with different initial orbital characteristics but with practically zero eccentricity. \emph{Top Left:} Eccentricity evolution of an EMRI with initial conditions $e(0)=0$, $p(0)/M=3.9$ and $\iota(0)=0.97$. \emph{Top Right:} Same as Top Left with $e(0)=0$, $p(0)/M=3$ and $\iota(0)=1.03$. \emph{Bottom Left:} Same as Top Left with $e(0)=0$, $p(0)/M=3.03$ and $\iota(0)=1.025$. \emph{Bottom Right:} Same as Top Right with $e(0)=0$, $p(0)/M=2.737$ and $\iota(0)=1.087$.}
    \label{eccentricity_integrable}
\end{figure*}

In Fig.~\ref{eccentricity_integrable}, we demonstrate four cases of resonant-crossing ($\Omega_r/\Omega_\theta=2$) EMRIs, where the primary has spin $a/M=0.85$ and the mass ratio of the secondary $\mu$ over the primary $M$ is fixed to $\mu/M=10^{-6}$, with different initial semi-latus rectum $p(0)/M$ and inclination $\iota(0)$. The eccentricity is initialized to be zero. The non-Kerr integrable metric of the central object has the following non-zero deformation parameters: $a_{12}=-1.5$, $a_{22}=4$, $a_{52}=6$ and $\epsilon_3=4$. Although we have made a meticulous scan of the parameter space $(a_{12},a_{22}, a_{52},\epsilon_3)$ in order to find a resonant maximization of eccentricity, we have observed that varying these deformation parameters by $\mathcal{O}(10^{-1})$ leads to qualitatively similar results.

On the top left of Fig. \ref{eccentricity_integrable} an integrable spherical EMRI is evolved, with initial zero eccentricity, through the resonance. We observe that at the time that is expected for the EMRI to cross the resonance, the eccentricity rapidly jumps to a non-zero value and the orbit remain eccentric till it crosses the separatrix. On the top right and bottom left of Fig. \ref{eccentricity_integrable} we observe a similar eccentricity excitation behavior with a twist. In the first case the trajectory is excited twice while in the second the orbit is first excited and later decays to a smaller but still non-zero eccentricity. Both cases have smaller semi-latus rectum and larger inclination, when the first resonance crossing takes place, compared with the top left of Fig. \ref{eccentricity_integrable}. In the former case the double excitation leads to a larger excitation of final eccentricity compared to all previously mentioned cases. Finally, by further decreasing $p/M$ and increasing $\iota$ at resonance crossing, we obtain a smoother transition to a non-spherical orbit while the eccentricity seems to reach its highest value (final eccentricity up to $\mathcal{O}(10^{-3})$).

\begin{figure}
    \includegraphics[scale=0.3]{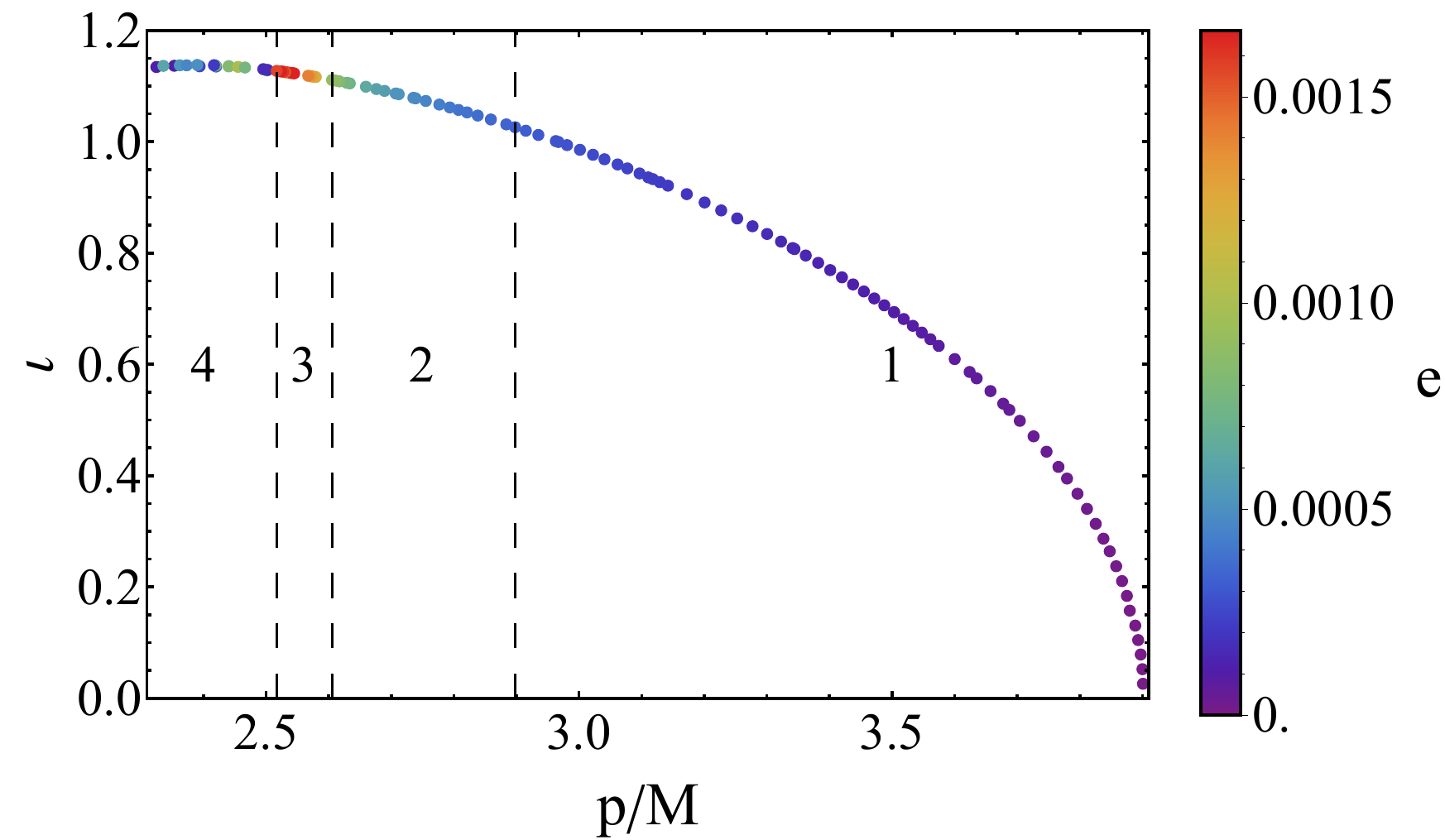}
    \caption{ Evolution of a spherical EMRI, where the supermassive primary is described by Eq.~\eqref{metric_tensor}, through the $\Omega_r/\Omega_\theta=2$ resonance. In all cases, we have $\mu/M=10^{-6}$, $a=0.85 M$, $a_{12}=-1.5$, $a_{22}=4$, $a_{52}=6$ and $\epsilon_3=4$. All points shown correspond to an initial eccentricity $e(0)=0$ and $\theta(0)=\pi/2$ that after the passage from the resonance obtain a constant non-zero value of eccentricity shown in the color bar. The points in region $1$ designate excitations of the EMRI's eccentricity with a single jump, as shown in the top left of Fig.~\ref{eccentricity_integrable}. The points in region $2$ and $4$ designate eccentricity excitations with double jumps/drops, as shown in the top right and bottom left of Fig.~\ref{eccentricity_integrable}. Finally, the points in region $3$ designate the initial conditions for spherical EMRIs that maximize the eccentricity excitation, as shown in the bottom right of Fig.~\ref{eccentricity_integrable}.}
    \label{rainbow_integrable}
\end{figure}

\begin{figure*}
    \includegraphics[scale=0.34]{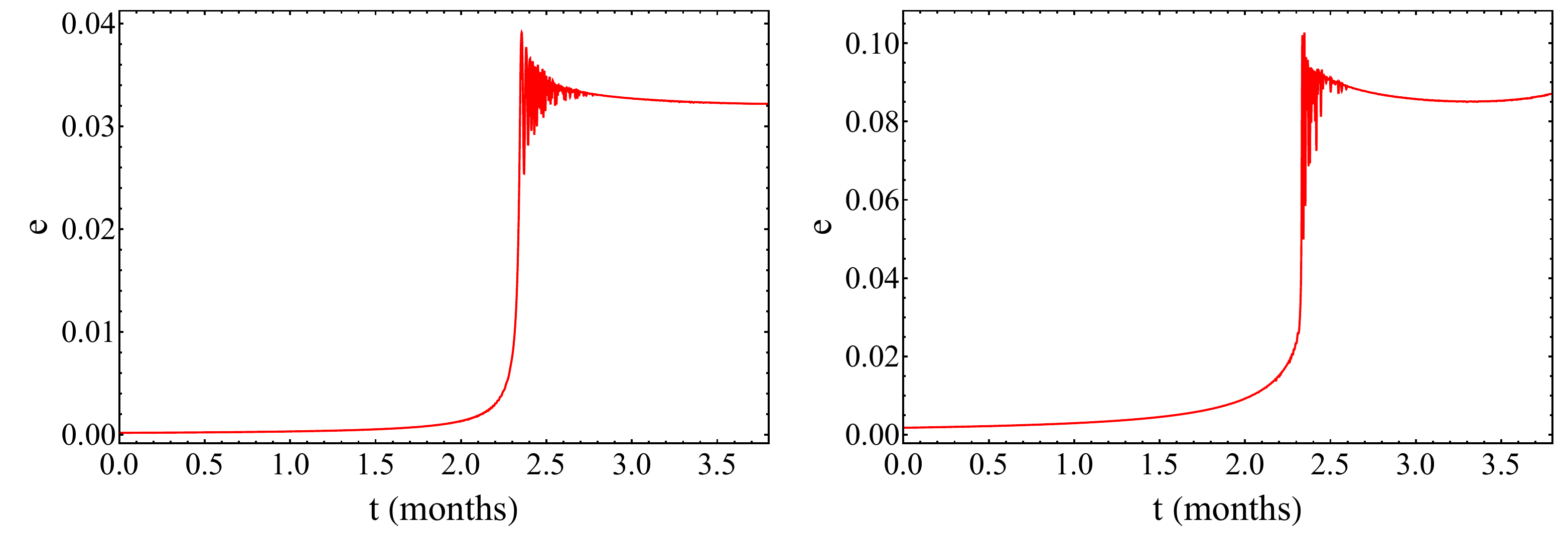}
    \caption{Evolution of a \emph{non-integrable}, spherical EMRI, where the supermassive primary is described by Eq. \eqref{metric_tensor}, through the $\Omega_r/\Omega_\theta=2$ resonance. In both cases, we have $\mu/M=10^{-6}$, $a=0.85 M$, $a_{12}=-1.5$, $a_{22}=4$, $a_{52}=6$, $\epsilon_3=4$ and non-zero $a_Q$. All orbits begin from the equatorial plane, i.e. $\theta(0)=\pi/2$, with different initial orbital characteristics but with practically zero eccentricity. \emph{Left:} Eccentricity evolution of an \emph{non-integrable EMRI} with initial conditions $r(0)/M=3.2$, $e(0)\sim0$, $E(0)/M=0.8098239$, $L_z(0)/M=0.7147996\mu$ and $a_Q=10^{-3}$. \emph{Right:} Same as Left with $a_Q=10^{-2}$.}
    \label{eccentricity_nonintegrable}
\end{figure*}

Intuitively, we could assume, from Fig. \ref{eccentricity_integrable}, that spherical, highly inclined, orbits that encounter the resonance  sufficiently close to plunge acquire the maximum eccentricity after they cross the resonance in study. What is really occurring though is hinted in Fig. \ref{eccentricity_integrable} when the resonant condition is met twice. In Fig. \ref{rainbow_integrable} we scan the parameter-space $(\iota, p/M)$ of spherical orbits that cross the resonance $\Omega_r/\Omega_\theta=2$. Each point corresponds to an initially spherical EMRI orbit that has been evolved to pass through the aforementioned resonance, starting with initial orbital parameters quite close to that of the resonance. The color map designates the value of eccentricity jump acquired after the passage through the resonance. The figure is divided into four regions to differentiate the various ways that eccentricity gets excited. Region 1 depicts initial conditions  that lead to a single jump in eccentricity when the resonance is crossed. These cases correspond to the one shown in Fig. \ref{eccentricity_integrable}, top left. Regions 2 and 4 correspond to double excitations of $e$ (either `double positive jumps' or `positive-negative jumps'), as shown in Fig. \ref{eccentricity_integrable}, top right and bottom left, respectively. In fact every point in region 2 leads to a subsequent point in region 4 since region 2 depicts the first jump in eccentricity while region 4 includes the positive/negative jumps of eccentricity at later times, when the orbit encounters the resonance for a second time. Interestingly, in region 3 we observe the largest eccentricity excitations with single prolonged jumps, as shown in the bottom right subfigure of Fig.~\ref{eccentricity_integrable}. These jumps occur similarly to the aforementioned case of double discrete passage through resonance, but now almost continuously, therefore they allow for degenerate positive jumps. Our parameter space search has shown that spherical inspirals around integrable non-Kerr BHs \emph{do not remain spherical}, since they finally acquire non-zero eccentricity due to the passage through the resonance $\Omega_r/\Omega_\theta=2$, in a large volume of the available parameter space. The phase-space Fig.~\ref{rainbow_integrable}, therefore, demonstrates that after resonance-crossing the eccentricity can grow up to order $\mathcal{O}(10^{-3})$. 

We point out that all our numerical integrations, regarding the adiabatic evolution of approximately spherical orbits, have been progressed up to times where the separatrix of the specific deformed-Kerr metric has not been crossed yet. Therefore, the observed increment in eccentricity is not an artifact that could be attributed to plunging behavior of the orbit, especially because it is too sharp. Furthermore, our numerical integration scheme remains valid all the way. In contrast, the orbital characteristics at the end-point of our integrations do lie in the area beyond the separatrix of Kerr, that is in the area describing plunging orbits of Kerr. Also we were careful in order to ensure that after the resonance crossing the eccentricity was ''stabilized'' for a period of time until it begins to increase again due to transition to plunge and we avoided orbits that after resonance crossing starts plunging almost immediately.

\section{Eccentricity excitation in non-integrable EMRIs}

In this section, we treat orbits of non-integrable EMRIs, that do not possess a Carter (or any other higher-rank Killing tensor) constant, thus integrability is broken and the binary can, in principle, demonstrate indirect imprints of chaos \cite{Contopoulos:2002,Destounis:2020kss}. This will then lead to a more pronounced resonant excitation of eccentricity in initially `spherical' EMRIs.

For the sake of argumentation we use the same, integrability-preserving, deformations as in the previous section, i.e. $a_{12}=-1.5$, $a_{22}=4$, $a_{52}=6$ and $\epsilon_3=4$, for an EMRI with a primary spin $a/M=0.85$ and mass ratio $\mu/M=10^{-6}$. On top of that setup, we turn on the deformation parameter $a_Q$ in order to observe its effect on resonant excitation of eccentricity. We note that when discussing initially `spherical'  orbits for a non-integrable EMRI ($a_Q\neq 0$), we do not commence with an exactly spherical orbit \emph{per se}. This is due to the fact that any integrability-breaking parameter of spacetime instantly renders the geodesic motion non-integrable, which in turn does not allow for the system to be expressed in action-angle variables. In what follows, we find initial conditions for resonant-crossing orbits with $\Omega_r/\Omega_\theta=2$ by assuming an integrable case with zero initial eccentricity in order to compute the energy, the angular momentum, and the initial semi-latus rectum, and then use these orbital characteristics for the non-integrable case. Thus our approximation of initial conditions works sufficiently well in order to observe the effect of eccentricity excitation in non-integrable EMRIs. Nevertheless, the absence of integrability induces small deviations from the initial conditions, with respect to those of the integrable system, and in particular in the initial eccentricity which is sufficiently small but non-zero, i.e. $e(0)\sim 10^{-4}-10^{-3}$ depending on the deformation parameter $a_Q$. Finally, we note that we have tested both positive and negative (small) values of $a_Q$ and the results were practically identical.

In Fig.~\ref{eccentricity_nonintegrable}, we demonstrate two representative inspirals with the same (deformation) properties as those in Fig.~\ref{eccentricity_integrable}, but with two discrete values of the integrability-breaking parameter added to the spacetime, i.e. $a_Q=10^{-3}$ (left subfigure) and $a_Q=10^{-2}$ (right subfigure). The initial non-zero eccentricities are $e(0)\sim 10^{-4}$ and $e(0)\sim 10^{-3}$, respectively, thus the initial minimal constraints of eccentricity seems to scale linearly with $a_Q$.  The results are fascinating in the sense that after the resonance-crossing, the practically spherical orbits acquire a rather notable final eccentricity of order $e\sim 10^{-2}$ (for $a_Q=10^{-3}$) and $e\sim 10^{-1}$ (for $a_Q=10^{-2}$), respectively Therefore, we expect a potentially-observable phenomenological effect when (non-)integrable spherical EMRIs are crossing previously uncharted parameter-space resonances as those with $\Omega_r/\Omega_\theta=2$.

In the previous sections we showed that, under radiation reaction, an integrable non-Kerr BH spacetime with a test-particle secondary inspiraling around it can cross the resonance $\Omega_r/\Omega_\theta=2$ when the deformation parameters are chosen randomly enough. These spherical EMRIs generically lead to a resonant amplification of eccentricity of initially spherical EMRIs and become eccentric. Furthermore, whenever we are able to additionally break integrability through $a_Q$, the initially (almost) spherical orbits evolve, after resonance-crossing, into eccentric inspirals with eccentricities of maximal order $\mathcal{O}(10^{-1})$. This is the first example of spherical orbits that under radiation reaction become non-spherical in non-Kerr EMRIs.

We note that the increment in eccentricity in the non-integrable case could not be attributed to the non-vanishing initial eccentricity, since our numerical investigation has shown that the final eccentricity after resonance-crossing in the integrable case is still of order $\mathcal{O}(10^{-3})$, even if the initial eccentricity is of the same order of magnitude with that encountered in the cases with non-zero $a_Q$ examined here.

\section{Gravitational-wave observables}

GWs carry information regarding the EMRI's fundamental frequencies, though in most cases these quantities could be deactivated through the action of previous evolution; i.e. the orbit has been circularized through radiation reaction. Here, we investigate the GW emission of the aforementioned demonstrative spherical EMRIs that undergo resonant excitations of eccentricity under radiation reaction. We will show that the frequency evolution of an incoming GW, detected by LISA, contain a clear imprint of eccentricity excitation when the $\Omega_r/\Omega_\theta=2$ resonance is present and is crossed by an initially spherical orbit.

To model GWs from EMRIs detected by the LISA interferometer we use the \emph{numerical kludge scheme} that combines exact particle trajectories with approximate GW radiation emission \cite{Babak:2006uv}. More sophisticated approaches involve Teukolsky-based waveforms deducing directly the GW characteristics through the Weyl scalars. Nevertheless, this method is quite intricate and does not elucidate further the main features presented here. The numerical kludge is perfectly-suited for phenomenology and has shown to agree extremely well ($\sim 95\%$) with Teukolsky-based waveforms \cite{Babak:2006uv}.

For this task, we will employ the quadrupole formula described below.

\subsection{Gravitational-wave modelling and LISA response}

The quadrupole formula takes advantage of the fact that the mass quadrupole emission of gravitational radiation is the dominant one, thus the radiative component of the metric perturbation introduced by a test particle at luminosity distance $d$ from the source $\boldsymbol{T}$ can be read at the transverse and traceless gauge as
\begin{equation}\label{metpert}
	h^\text{TT}_{ij}=\frac{2}{d}\frac{d^2 I_{ij}}{dt^2},
\end{equation}
where $I_{ij}$ is the symmetric and trace-free (STF) mass quadrupole tensor
\begin{equation}
	I^{ij}=\left[\int x^i x^j T^{tt}(t,x^i) \,d^3 x\right]^\text{STF},
\end{equation}
with $t$ being the coordinate time measured at infinite distance from the source. The source term of the secondary is
\begin{equation}\label{Ttt}
	T^{tt}(t,x^i)=\mu \delta^{(3)}\left[x^i-Z^i(t)\right],
\end{equation}
where $Z(t)=(x(t),y(t),z(t))$, with 
\begin{align}
    x(t)&=r(t)\sin\theta(t)\cos\phi(t),\\
    y(t)&=r(t)\sin\theta(t)\sin\phi(t),\\
    z(t)&=r(t)\cos\theta(t),
\end{align} 
includes the components of the orbit with respect to flat spherical coordinates, under the assumption that our detector is placed at infinity. Here, we identify the Schwarzschild coordinates $(r,\theta,\phi)$ of the secondary trajectory with Minkowskian coordinates; a methodology known as the `particle-on-a-string' approximation, since we assume a finite luminosity distance $d$ from the source. Even though this technique is not strictly valid, it has been found to work very well when generating EMRI waveforms \cite{Babak:2006uv}.

The incoming GWs onto the space-borne detector can be projected on their two polarizations, $+$ and $\times$, by introducing two unit vectors, $\boldsymbol{p}$ and $\boldsymbol{q}$, which are defined with respect to a third unit vector $\boldsymbol{n}$ that points from the EMRI to the detector. This triplet of unit vectors $\boldsymbol{p},\,\boldsymbol{q},\,\boldsymbol{n}$ is chosen so that they form an orthonormal basis. The polarization tensor components are then given by
\begin{equation}
	\epsilon_+^{ij}=p^i p^j-q^i q^j,\,\,\,\,\,\,\epsilon_\times^{ij}=p^i q^j+p^j q^i.
\end{equation}
The polarization tensor allow us to write the metric perturbation as
\begin{equation}
	h^{ij}(t)=\epsilon_+^{ij}h_+(t)+\epsilon_\times^{ij}h_\times(t),
\end{equation}
where
\begin{equation}
	h_+(t)=\frac{1}{2}\epsilon_+^{ij}h_{ij}(t),\,\,\,\,\,\,\,h_\times(t)=\frac{1}{2}\epsilon_\times^{ij}h_{ij}(t).
\end{equation}
Then, one can express the GW components $h_{+,\times}(t)$ in terms of the position vector $Z^i(t)$, the velocity vector $v^i(t)=dZ^i/dt$, and the acceleration vector $a^i(t)=d^2Z^i/dt^2$, as
\begin{equation}
	\label{GW_formula}
	h_{+,\times}(t)=\frac{2\mu}{d}\epsilon^{+,\times}_{ij}\left[a^i(t)Z^j(t)+v^i(t)v^j(t)\right].
\end{equation}

LISA's response to an incident GW depends on the antennae patterns of the detector $F^{+,\times}_{I,II}$ (see Refs. \cite{Cutler:1997ta,Barack:2003fp,Destounis:2020kss} for their exact functional forms), therefore the total gravitational waveform detected by LISA is
\begin{equation}\label{total_GW}
	h_\alpha(t)=\frac{\sqrt{3}}{2}\left[F^+_\alpha(t) h_+(t)+F^\times_\alpha(t) h_\times(t)\right],
\end{equation}
where $\alpha=\{I,II\}$ are the channel indices of the detector's antennae. Our analysis is simplified by the assumption that LISA lies at a luminosity distance $d$ with fixed orientation $\boldsymbol{n}=(0,0,1)$ with respect to the source and that the primary's polar and azimuthal angles are fixed at the equatorial plane (this choice simplifies significantly the antennae response patterns). 
	
A generic data stream captured by LISA contains both the clear signal of the source together with some noise. In our case we assume that the noise is stationary and Gaussian with zero mean. We also assume that the two data streams sectors are uncorrelated and the noise power spectral density of LISA $S_n(f)$ (that includes instrumental, galactic and extra-galactic confusion noise \cite{Cutler:1997ta,Barack:2003fp}) is equivalent at both channels. This allows for a single-channel approximation which we will employ in what follows (see Refs. \cite{Cutler:1997ta,Barack:2003fp,Canizares:2012is,Destounis:2020kss} for more details).

\subsection{Fourier analysis and spectrograms}

Equation \eqref{total_GW} provides an accurate approximation of the GWs detected by LISA from the radiation emitted by an EMRI. Gravitational waveforms are captured in the time domain. Nevertheless, there exist a handful of data analysis and signal processing techniques that can maximize the phenomenological yield from GW observations.

\begin{figure}
    \includegraphics[scale=0.31]{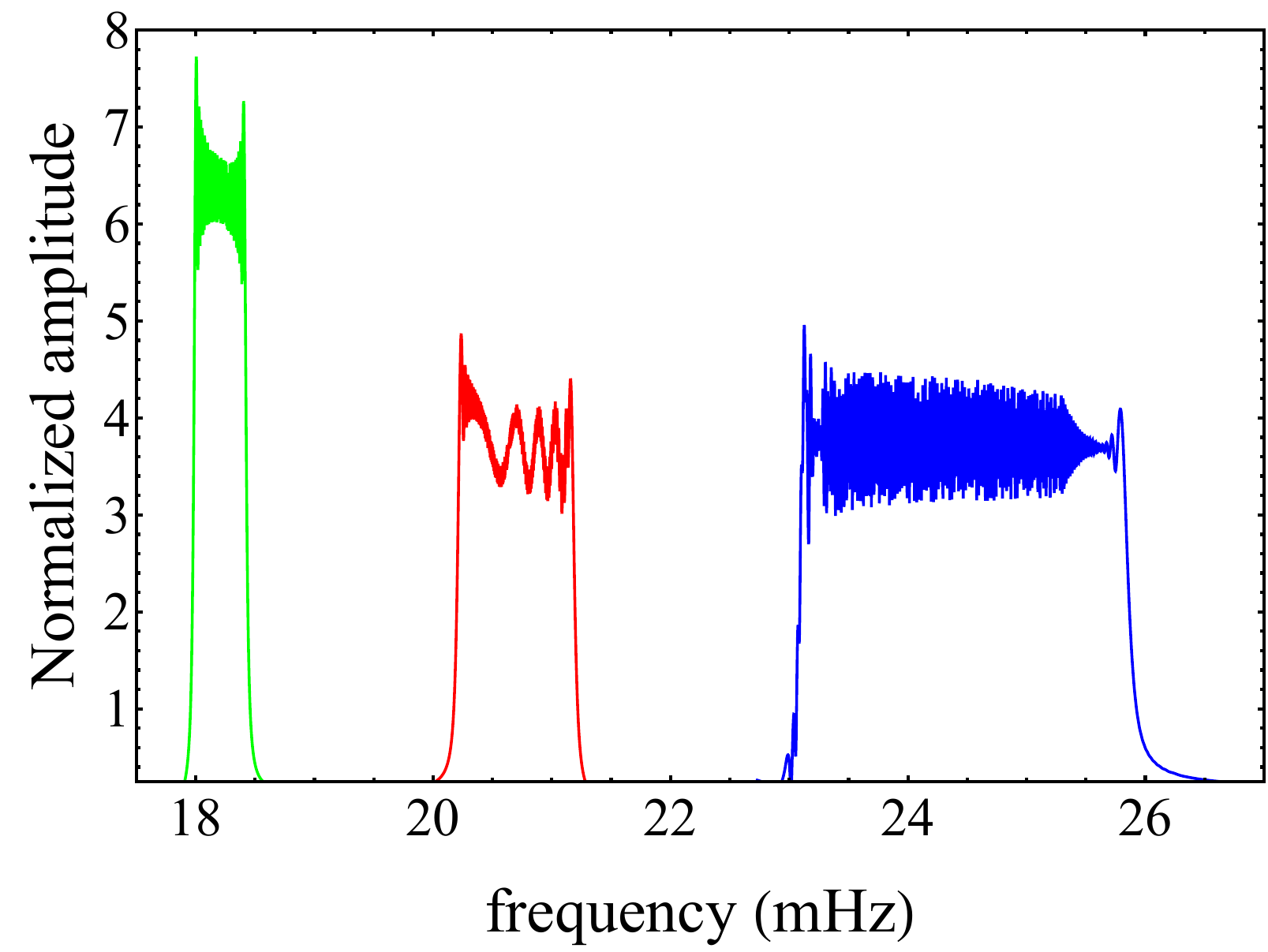}
    \caption{Short-time Fourier transform of three different time segments of the fundamental harmonic of the GW emitted by the integrable EMRI shown in Fig. \ref{eccentricity_integrable}, bottom right. Different colors designate the Fourier transform of different time segments of the GW signal, i.e. before (green), during (red) and after (blue) the eccentricity excitation.}
    \label{STFT_integrable}
\end{figure}

One of the main tools in signal processing and data analysis is the Fourier transform of the time-domain signal onto the frequency domain. This helps unravel underlying phenomenology that is hard (if not impossible) to spot in time-domain signals. In what follows, we assume time-domain waveforms $h(t)$ and frequency-domain ones, after being Fourier transformed, as $\tilde{h}(f)$, where $f$ is the frequency. A Fourier-transformed signal is represented with imaginary numbers, therefore we usually take its absolute value in order to present Fourier peaks, and thus the resulting GW spectrum of the EMRI. The Fourier transform convention we assume is
\begin{equation}
	\tilde{h}(f)=\int_{-\infty}^{\infty}e^{i2\pi f t}h(t)dt.
\end{equation}

\begin{figure}
    \includegraphics[scale=0.3]{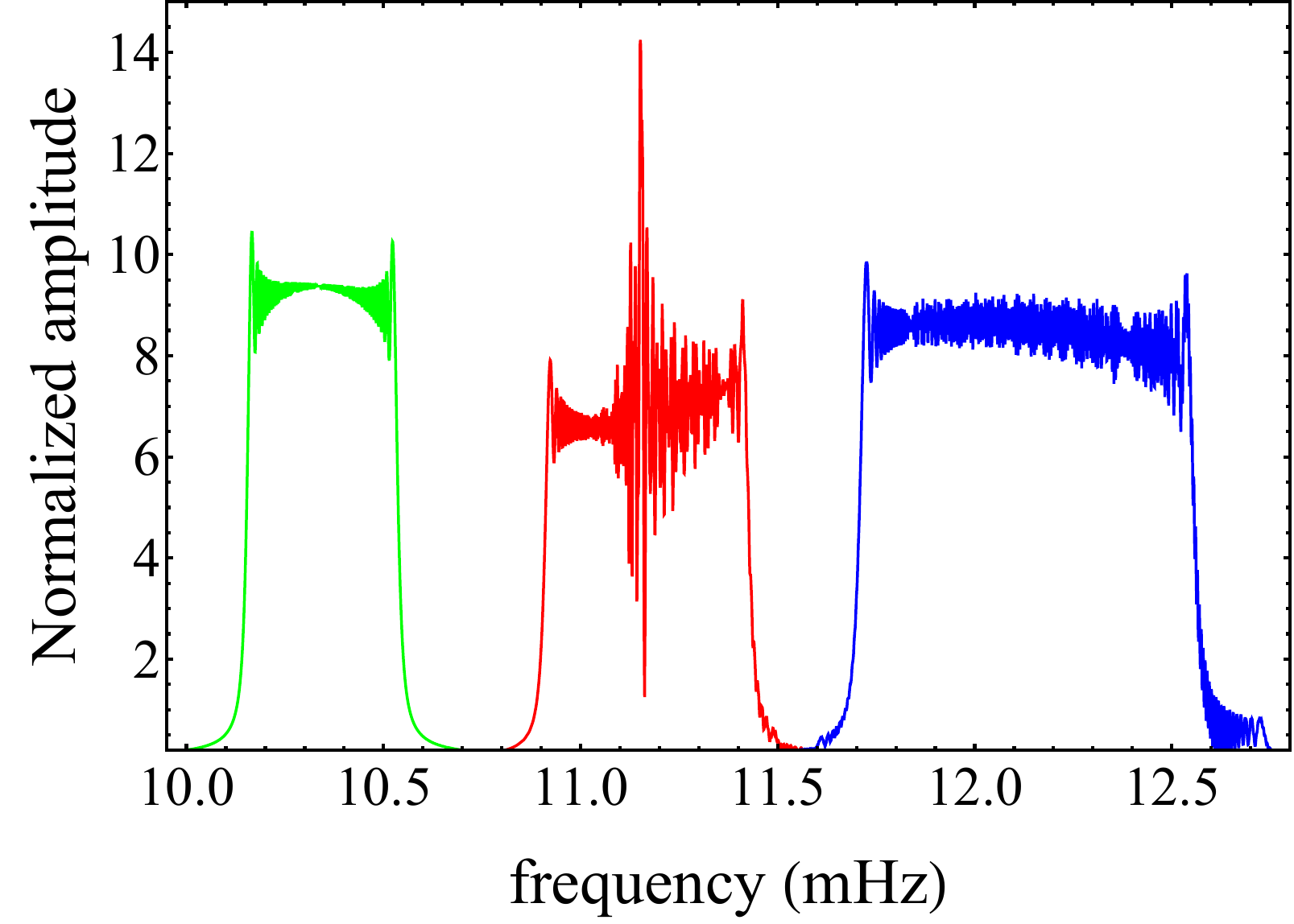}
    \caption{Short-time Fourier transform of three different time segments of the fundamental harmonic of the GW emitted by the non-integrable EMRI shown in Fig. \ref{eccentricity_nonintegrable}, right, where $a_Q=10^{-2}$. Different colors designate the Fourier transform of different time segments of the GW signal, i.e. before (green), during (red) and after (blue) the eccentricity excitation.}
    \label{STFT_nonintegrable}
\end{figure}

Another method to study GW frequency dynamics is through spectrograms. To construct a spectrogram we use consecutive short-time Fourier transforms (STFTs); a well-known method used to determine the frequency content of local time segments of a signal as it changes over time. By dividing the waveform into short time segments of equal length, we calculate consecutive Fourier transforms on each segment. Then the evolution of GW frequencies from one temporal segment to another is plotted, known as a spectrogram (frequency vs time plot). To achieve a smoother transition between segments, we employ a window of fixed size with an offset, which we slide over the available signal. For each such window the Fourier transform is computed. This method allows overlapping of time segments and leads to smoother frequency transitions in spectrograms. The choice of the window size is related to an uncertainty principle as the standard deviation in frequency and time is limited in these figures. In what follows, we employ a window with an appropriate offset so that we represent the frequency evolution of the given signals as optimal as possible.

\subsection{Gravitational-waves from integrable EMRIs}

Following the methodologies of the previous subsections we obtain approximate gravitational waveforms detected by LISA from the exact secondary trajectory of resonant-crossing, integrable EMRI orbits that are initially spherical. For the sake of argumentation, here we will produce the waveform of the orbit designated in Fig.~\ref{eccentricity_integrable}, bottom right, since it produces qualitatively the largest resonant excitation of the EMRI's eccentricity from zero to $\mathcal{O}(10^{-3})$.

In Fig.~\ref{STFT_integrable} we have chosen three distinct time segments of the produced GW signal, i.e. one where the eccentricity is still zero (green peak), another at the moment of eccentricity growth driven by resonance crossing (red peak) and a final one after the resonance has been crossed, where the final EMRI is eccentric (blue peak). We perform a Fourier transform on all three time segments and present the fundamental harmonic, i.e. the one that has the largest initial amplitude, as it evolves in the frequency domain. Firstly, we observe that as we move from one temporal segment to the next an obvious `stretching' of frequency range occurs. This is due to the fact that as time evolves the secondary orbital period shrinks and the overall evolution of the EMRI is accelerated due to radiation reaction. Thus, consecutive time segments (from left to right in Fig. \ref{STFT_integrable}) span successively larger frequency ranges. The most important observation one can draw is the fact that the middle (red) peak, that represents a time segment where the resonance crossing occurs, is oscillating with two different frequencies; the initial highly-oscillatory harmonic that appears even in the green peak plus a secondary frequency attributed to the resonance crossing. The secondary frequency, that is initially not excited, gains amplitude during resonance-crossing through the modulation of the fundamental frequency's magnitude, where parts of this frequency line are pronounced while others are suppressed. This phenomenon might designate an energy transfer mechanism between the fundamental frequencies of the orbit. In any case, the resonantly-excited frequency is of order $\mathcal{O}(1)$ in normalized amplitude, therefore it might be distinguishable by spaceborne detectors. It is noteworthy to state that after the crossing of the resonance other peaks (that are not shown in Fig.~\ref{STFT_integrable}) are excited as well, but their amplitudes are of order $\mathcal{O}(10^{-2})$, thus we do not depict them. We will see though, that the effect of integrability-breaking may lead to potential new GW observables.

\subsection{Gravitational-waves from non-integrable EMRIs}

While integrable deformations of Kerr EMRIs allow for the breaking of sphericity in initially-spherical orbits under radiation reaction, as it is possible for the resonance $\Omega_r/\Omega_\theta=2$ to be crossed, we can ascertain that not all rotating compact object EMRIs that are initially spherical will remain spherical under radiation reaction. This result serves as a `proof-of-principle' since in the previous subsection we established that the such effect might be observable with future space-borne detectors. In this subsection, we find even stronger observables that occur due to the excitation of eccentricity in non-integrable EMRIs.

We obtain approximate gravitational waveforms detected by LISA from the secondary's trajectory orbiting a non-integrable EMRI that crosses a resonance as before. We produce the waveform of the orbit designated in Fig.~\ref{eccentricity_nonintegrable}, right panel, which gives the strongest `resonant kick' in eccentricity, of order $\mathcal{O}(10^{-1})$, when $\Omega_r/\Omega_\theta=2$ is satisfied. 

\begin{figure}[t]
    \includegraphics[scale=0.3]{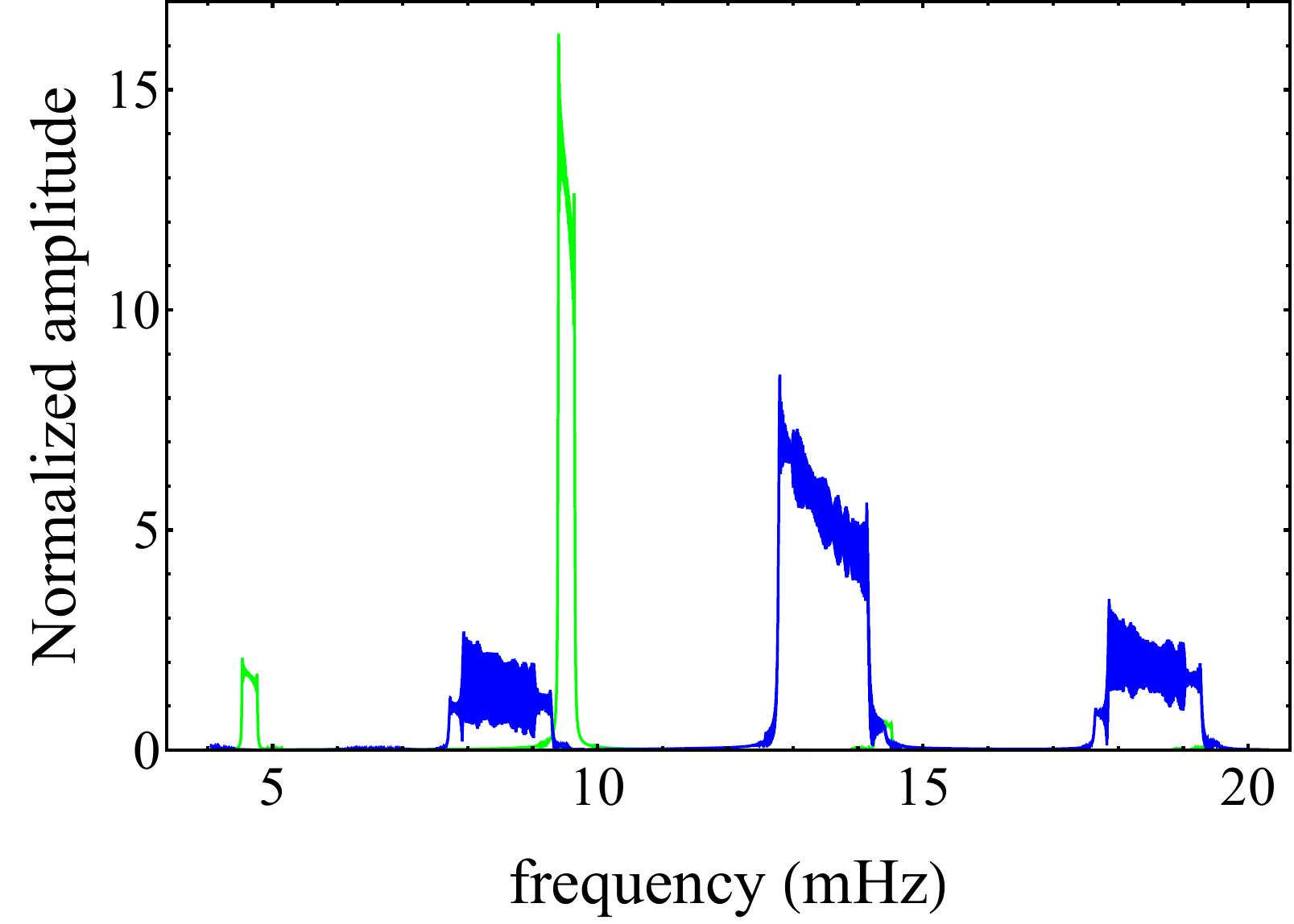}
    \caption{Short-time Fourier transform of two different time segments of the fundamental, and the two higher harmonics, of the GW emitted by the \emph{non-integrable EMRI} shown in Fig.~\ref{eccentricity_nonintegrable}, right, with $a_Q=10^{-2}$. Different colors designate the Fourier transform of different time segments of the GW signal, before (green) and after (blue) the eccentricity excitation. It is evident that two harmonics gain amplitude after the resonant excitation.}
    \label{STFT_nonintegrable_zoom_out}
\end{figure}

\begin{figure*}
    \includegraphics[scale=0.65]{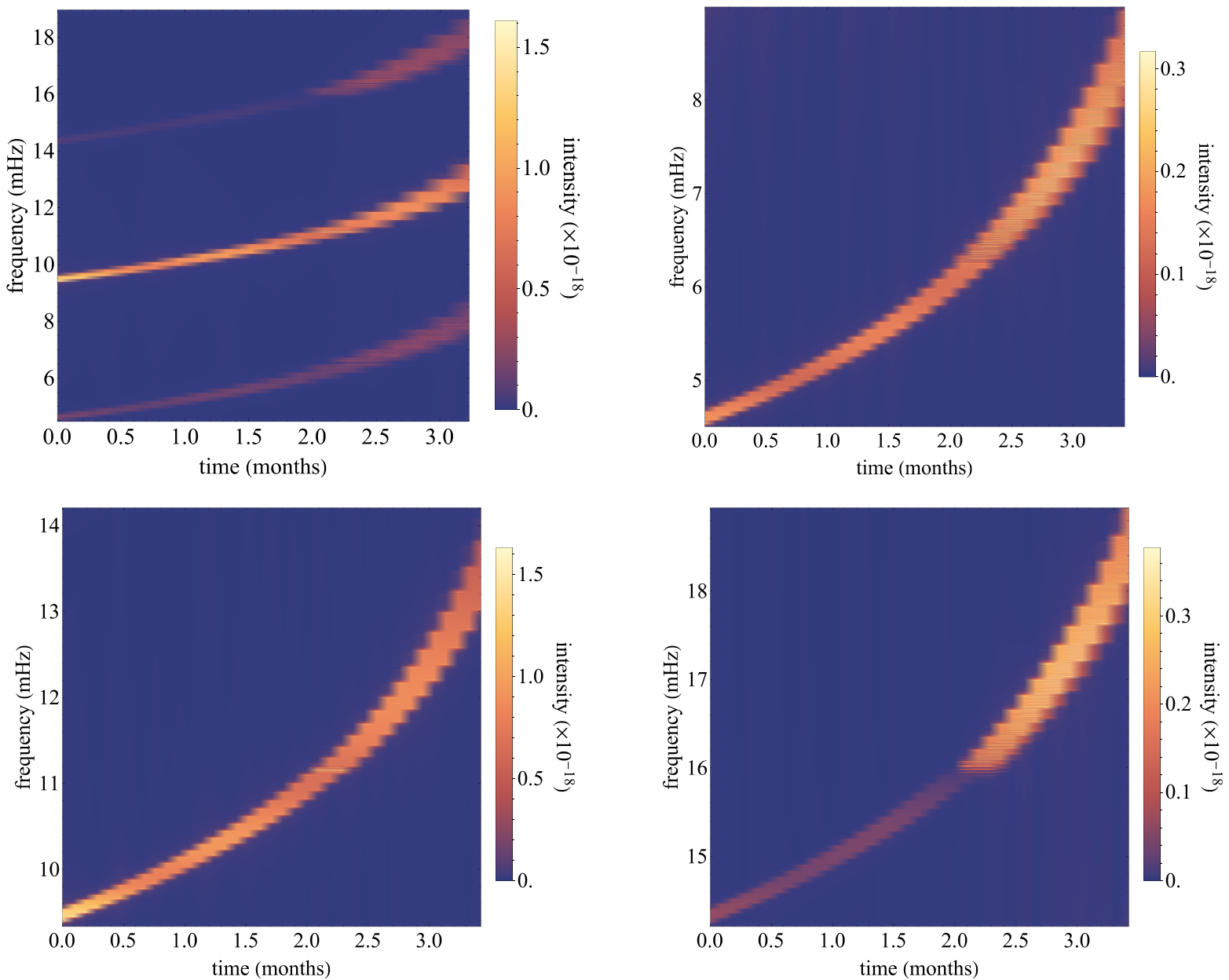}
    \caption{\emph{Top left:} Spectrogram of the three most dominant harmonics of the GW emitted from the \emph{non-integrable EMRI} in Fig. \ref{eccentricity_nonintegrable}, right. \emph{Top right:} Spectrogram of the low-frequency harmonic (see Top left) of the GW emitted from the \emph{non-integrable EMRI} in Fig. \ref{eccentricity_nonintegrable}, right. \emph{Bottom left:} Same as Top right but for the most dominant harmonic (see Top left) of the GW emitted from the \emph{non-integrable EMRI} in Fig. \ref{eccentricity_nonintegrable}, right. \emph{Bottom right:} Same as Top right but for the high-frequency harmonic (see Top left) of the GW emitted from the \emph{non-integrable EMRI} in Fig. \ref{eccentricity_nonintegrable}, right.}
    \label{Spectrogram_nonintegrable}
\end{figure*}

In Fig.~\ref{STFT_nonintegrable} we have chosen again three time segments of the produced GW signal with $a_Q=10^{-2}$, i.e. before (green peak), during (red peak) and after (blue peak) the resonance crossing. These have been Fourier-transformed for their normalized amplitude to be presented with respect to the GW frequency. The most notable observation we make from Fig.~\ref{STFT_nonintegrable} regards the red peak which, as before, is the one that depicts the time segment during the resonance crossing. Interestingly, the particular peak, again, includes two frequencies, the pre-resonant-crossing one that appears before and after the eccentricity excitation and another frequency that is excited during resonance-crossing and leads to an amplitude modulation of order $\mathcal{O}(10^1)$. This notable frequency excitation is an attribute of the rather profound eccentricity excitation when integrability is broken, which stems from the fact that resonances in non-integrable systems `grow' resonant islands around them which prolong the periodicity of the otherwise zero-volume resonant points in integrable systems (for more information see \cite{Contopoulos:2002,Apostolatos:2009vu,Destounis:2021mqv,Destounis:2021rko}).

To further elaborate on the phenomenon of eccentricity excitation when a resonant island is met in initially spherical non-integrable EMRIs, we focus on the first three harmonics of two different time segments: one before crossing the resonance and another after the resonance crossing and eccentricity excitation has occurred. In Fig.~\ref{STFT_nonintegrable_zoom_out} we show the three most dominant Fourier peaks before crossing the resonant island (green peaks). We observe that besides the highest peak that resides around $10$ mHz, there exist another peak close to $4$ mHz, and another close to $14$ mHz that is practically indistinguishable. These small-amplitude harmonics exist due to the tiny, though non-zero, initial eccentricity of the non-integrable EMRI, as discussed in previous sections. With blue (in the same figure), we show the same three harmonics that have evolved through the resonant island and the orbit acquired an eccentricity of order $\mathcal{O}(10^{-1})$. We observe that after the resonance all three peaks modulate at the same order of magnitude and in particular the excited (blue) peak around $18$ mHz has gained amplitude of at least $\mathcal{O}(10^1)$ with respect to the corresponding rightmost green peak. This GW frequency amplification is attributed to the non-zero eccentricity gained after crossing the resonance.

To complete our illustration of the particular phenomenological GW observable of spherical-to-eccentric orbit transition, we sketch the spectrograms (frequency evolution) of the three dominant peaks from Fig. \ref{STFT_nonintegrable_zoom_out} in green with respect to coordinate time $t$. In Fig. \ref{Spectrogram_nonintegrable} (top left) we demonstrate clearly the phenomenon of eccentricity excitation at the orbital level that leads to the excitation and amplitude modulation of subharmonics after the EMRI crosses the resonance. We observe the evolution of three frequencies; the central one that begins evolving at around $10$ mHz, together with two subharmonics that begin evolving at around $4$ (leftmost peak shown in Fig. \ref{STFT_nonintegrable_zoom_out}) and $14$ mHz (rightmost peak shown in Fig. \ref{STFT_nonintegrable_zoom_out}), respectively. It is evident that the magnitude of the central frequency is modulated; its intensity after resonance-crossing, at around $2.1$ months of inspiral, decreases (see also bottom left in Fig. \ref{Spectrogram_nonintegrable}). At the same time both subharmonics gain amplitude after crossing the resonance (see top and bottom right in Fig. \ref{Spectrogram_nonintegrable}); an effect that clearly demonstrates the frequency modulation due to the resonant excitation of eccentricity. We notice that the two subharmonics have non-zero intensity. This is due to the fact that the non-integrable EMRI is not initially exactly spherical but rather slightly eccentric, therefore these frequencies are already moderately visible in the spectrogram. Specifically, the highest frequency in Fig. \ref{Spectrogram_nonintegrable}, top left panel, presents a quite visible, order of magnitude, intensity growth after resonance (see also Fig. \ref{Spectrogram_nonintegrable}, bottom right), while the lowest frequency has less visible intensity gain. Finally, the central frequency undergoes a visible, order of magnitude, intensity drop after resonance with an order of magnitude jump appearing right at the resonant orbit. This rapid intensity gain corresponds to the large oscillation present in the red peak of Fig. \ref{STFT_nonintegrable}. Hence, if future space-detectors are accurate enough then such simplistic, though highly interesting, phenomenon should appear in their frequency band.

\section{Conclusions}

In \cite{Apostolatos:1993nu,Kennefick:1995za}, it has been shown that spherical EMRIs with Schwarzschild and Kerr primaries remain spherical (zero eccentricity) under radiation reaction. Here, we have studied initially-spherical orbits of rotating, non-Kerr primary BHs in EMRIs, under radiation reaction, in order to test if spherical orbits always remain spherical, even when the primary is not a Kerr BH. 

We first, found that non-Kerr compact object, such as those described by the Johanssen metric, provide an enlarged volume of phase-space, with respect to that of Kerr EMRIs, and extends to strong-field regions of trajectories where resonant orbits with multiplicity $\Omega_r/\Omega_\theta=2$ take place, in contrast to Kerr orbits that do not have access to this resonances. The transversal of such resonances leads to a variety of phenomenology since the EMRI, in principle, will gain eccentricity and eventually demonstrate potentially-observable imprints in the GW band of LISA.

When evolving the small secondary companion around integrable non-Kerr EMRIs, e.g. the Johanssen primary \cite{Johannsen} that still exhibits a Carter-like constant, we observed a small but rather important increment of eccentricity from zero to $\mathcal{O}(10^{-3})$ when the resonance condition is met. This phenomenon might be, in principle, observable by space detectors. Nevertheless, we found that there are ways to amplify the resonant excitation of eccentricity up to $\mathcal{O}(10^{-1})$. Indeed, we have shown that there are cases where the integrability of geodesics is \emph{slightly} broken and the effect of resonant excitation of eccentricity becomes more profound.

In particular, non-Kerr EMRIs, with non-integrable geodesics, have been evolved in the same setup. We found that due to the growth of a resonant island around the resonance $\Omega_r/\Omega_\theta=2$, though small, is influential enough to excite the eccentricity of initially-spherical EMRIs (with $e=0$) to orders of $\mathcal{O}(10^{-1})$. Even though the integrable inspiral crossings of the resonant (periodic) orbit does not induce such a substantial change in eccentricity, non-integrable, resonant-crossing EMRIs are substantial laboratories for inspirals that begin their evolution with zero eccentricity until they advance into eccentric orbits with eccentricity of order $\mathcal{O}(10^{-1})$. The advance of non-integrable, initially-spherical EMRIs, into eccentric trajectories is imprinted onto the GWs detected by such sources through an amplitude modulation of their frequencies. Remarkably, even when the GW spectrum of the EMRI is initially (almost) monochromatic, due to the sphericity of the secondary's trajectory, we find that the final, resonantly-enhanced EMRI contains an oligochromatic GW spectrum with excited GW frequencies related to the eccentric nature of the final EMRI. These new `voices' in the GW spectrum may serve as a smoking-gun for the non-Kerrness of the supermassive primary.

Even though GW signals are described by radial, polar and azimuthal voices the dynamics between individual GW voices can be made to dominate by varying the eccentricity and inclination \cite{Drasco:2005kz}. Although each voice is generally apparent in the EMRI waveform, the radial one is prone to overpowering the others. Thus, in our case which describes a spherical-to-generic orbital transition, due to resonant effects, we should be able to distinguish between monochromatic and oligochromatic GW frequency spectra of EMRIs by focusing on their radial transitions, such as the evolution of eccentricity \cite{Seoane:2024nus}. Since  each voice evolves in a simple way on long timescales, one can exploit such property to efficiently produce waveform models that faithfully encode the properties of EMRI systems \cite{Hughes:2021exa} and in particular the numerous effects of transient orbital resonances. 

\begin{acknowledgments}
The authors acknowledge financial support provided by the DAAD program for the “Promotion of the exchange and scientific cooperation between Greece and Germany IKYDAAD 2022” (57628320). This project has received funding
from the European Union’s Horizon-MSCA-2022 research and innovation programme ``EinsteinWaves'' under grant agreement No. 101131233. K.D. acknowledges financial support provided under the European Union’s H2020 ERC, Starting Grant agreement no. DarkGRA–757480 and the MIUR PRIN and FARE programmes (GW-NEXT, CUP:B84I20000100001). K.D. also acknowledges hospitality  from the National and Kapodistrian University of Athens, and University of T\"ubingen where part of this work was conducted.
\end{acknowledgments}

\begin{appendix}
\section{Constants of motion}\label{App. A}

In Appendices \ref{App. A} and \ref{App.B} we have used the dimensionless quantities:
$\tilde{E}=\frac{E}{\mu}$, $\tilde{L_z}=\frac{L_z}{\mu M}$, $\tilde{\tilde{Q}}=\frac{\tilde{Q}}{\mu^2 M^2}$, $\tilde{r}=\frac{r}{M}$,
$\tilde{a}=\frac{a}{M}$ and $\tilde{p}=\frac{p}{M}$.
So, we can assume $M=\mu=1$ and keep working with all the above reduced quantities without tildes. 

For the integrable system $a_Q=0$ the constants of motion $\{E, L_z, \tilde{Q}\}$ can be expressed in terms of the orbital parameters: the semi-latus rectum $p$, the eccentricity $e$ and the minimum polar angle $\theta_{min}$ reached by the orbit. We derive the corresponding expressions for the constants of motion in Johannsen spacetime following the method outlined in Appendix B of \cite{2002CQGra..19.2743S} for the Kerr metric.

The condition of polar turning points, i.e. $\Theta(\theta_{\rm min})=0$, can be used to express $\tilde{Q}$ as a function of $(E,L_z,z_-=\cos^2{\theta_{\rm min}})$ as
\begin{equation}\label{Qcomp}
    \tilde{Q}=z_{-}\left[
    a^2(1-E^2)+\frac{L_z^2}{1-z_{-}}
    \right].
\end{equation}
Using Eq. (\ref{Qcomp}), the radial potential (\ref{R}) can be written as
\begin{equation}
    R(r)=f(r)E^2-2g(r)E L_z-h(r)L_z^2-d(r),
\end{equation}
where
\begin{align*}
f(r)&=(r^2+a^2)^2A_1^2(r)-a^2(1-z_{-})\Delta,\\
g(r)&=a(r^2+a^2)A_1(r)A_2(r)-a\Delta,\\
h(r)&=-a^2A_2^2(r)+\frac{\Delta}{1-z_{-}},\\
d(r)&=\Delta[r^2+f(r)+z_{-}a^2].
\end{align*}
Solving the system of equations $R(r_1)=R(r_2)=0$, and for spherical orbits ($r_1=r_2=r_0$) the equations $R(r_0)=0$ and $R'(r_0)=\frac{dR(r)}{dr}|_{r_0}=0$, the energy $E$ is given by
\begin{equation}\label{Ecomp}
E^2=\frac{\kappa\rho+2\epsilon\sigma- 2D \sqrt{\sigma(\sigma\epsilon^2+\rho\kappa\epsilon-\eta\kappa^2)}}{\rho^2+4 \eta\sigma},
\end{equation}
where $D=\pm 1$ and the determinants $\kappa, \rho, \sigma, \epsilon, \eta$
are defined as
\begin{eqnarray*}\label{dets1}
\kappa&=d_1h_2-d_2h_1,\\
\rho&=f_1h_2-f_2h_1,\\
\sigma&=g_1h_2-g_2h_1,\\
\epsilon&=d_1g_2-d_2g_1,\\
\eta&=f_1g_2-f_2g_1.
\end{eqnarray*}
In the above expressions for the determinants the subscripts $1,\, 2$ have the following meaning:
\begin{enumerate}
\item[i.] for eccentric orbits ($e\neq 0$):
\begin{eqnarray} \label{fgd0}
(f_1,g_1,h_1,d_1)&=(f(r_1),g(r_1),h(r_1),d(r_1)),\\
(f_2,g_2,h_2,d_2)&=(f(r_2),g(r_2),h(r_2),d(r_2)),
\end{eqnarray}
\item[ii.] for spherical orbits ($e=0$ and $r_1=r_2=r_0$):
\begin{eqnarray} \label{fgdn0}
(f_1,g_1,h_1,d_1)&=(f(r_0),g(r_0),h(r_0),d(r_0)),\\
(f_2,g_2,h_2,d_2)&=(f'(r_0),g'(r_0),h'(r_0),d'(r_0)).
\end{eqnarray}
\end{enumerate}
Finally, the angular momentum is given by:
\begin{equation}
    \label{Lzcomp}
L_z=-\frac{g_1E}{h_1}+\frac{D}{h_1}\sqrt{g_1^2E^2+(f_1E^2-d_1)h_1},
\end{equation}
where $D=1$ refers to prograde and $D=-1$ to retrograde orbits.

\section{Radial integrals in \boldmath\texorpdfstring{$\chi-$}{x}representation}\label{App.B}

In this Appendix we derive the expressions for the radial integrals (\ref{Y})-(\ref{X}), under the substitution:
\begin{equation}\label{chi}
    r(\chi)=\frac{1}{\lambda(\chi)}  \quad \mbox{where} \quad \lambda(\chi) =\frac{1+e\cos{\chi}}{p}\, .
\end{equation}

The radial potential (\ref{R}) is rewritten as: 
\begin{equation}\label{Rx}
    R(r)=\frac{V_{10}(r)}{r^6}, \qquad V_{10}=(r_1-r)(r-r_2) \prod_{i=3}^{10} (r-r_i),
\end{equation}
where $V_{10}(r)$ is a polynomial of tenth degree and $r_i$ with $i=1,2,..,10$ are the ten real and complex roots of $R(r)$.

By analysing the expression (\ref{Rx}) and equating to (\ref{R}) we can relate the coefficients of $r^n$, $n=1,..,10$, with sums of products of the roots $r_i$.

Therefore the polynomial $V_{10}$ using (\ref{chi}) becomes:
\begin{equation*}
    V_{10}(\chi)=\frac{1}{\lambda(\chi)^{10}}\frac{e^2\sin^2{\chi}}{(1-e^2)^2}J_e(\chi).
\end{equation*}
While, the radial integrals in Eqs. (\ref{Y})-(\ref{X}) in the $\chi-$representation become:
\begin{align}
    Y & =  p (1-e^2) \int_0^{\pi}\frac{1+P(\chi)}{{\tilde \lambda(\chi)}^2{\cal P}(\chi)}d\chi,\\
    W & =  p(1-e^2)\int_0^{\pi}\frac{F(\chi)+E F_1(\chi)-a L_z /r(\chi)^2F_2(\chi)}{{\tilde \lambda(\chi)}^2H(\chi){\cal P}(\chi)}d\chi,\\
    Z & = \frac{1-e^2}{ p}\int_0^{\pi}\frac{G(\chi)+a E G_1(\chi)
    -a^2 L_z/r(\chi)^2G_2(\chi)}{H(\chi){\cal P}(\chi)}d\chi,\\
    X & =  \frac{1-e^2}{p}\int_0^{\pi}\frac{d\chi}{{\cal P}(\chi)},
\end{align}
where ${\tilde \lambda(\chi)}\equiv\lambda(\chi) p \equiv p/r(\chi) $ and the functions $A_5(\chi)$, $P(\chi)$, $F(\chi)$, $F_1(\chi)$, $F_2(\chi)=G_1(\chi)$, $G(\chi)$, $G_2(\chi)$, $H(\chi)$ and ${\cal P}(\chi)$ are defined by:
\begin{align*}
A_5(\chi)& = 1+a_{52}\lambda(\chi)^2\, ,\\
P(\chi)& = \epsilon_3\lambda(\chi)^3\, ,\\
F(\chi)& = \left[1+a^2\lambda(\chi)^2\right]E
- 2 a \lambda(\chi)^3 (L_z-a E)\, ,\\
 F_1(\chi)& = a_{13}\lambda(\chi)^3 
\left[1+a^2\lambda(\chi)^2\right]^2\left[2+a_{13}\lambda(\chi)^3\right]\, ,\\
 F_2(\chi)& =
\lambda(\chi)^2\left[1+a^2 \lambda(\chi)^2\right]  \left[a_{22}+\right.\\
&\left. +a_{13}\lambda(\chi) 
\left(1 +a_{22}\lambda(\chi)^2\right)\right]\, ,
\\
 G(\chi)& = L_z-2\lambda(\chi)(L_z-a E)\, , \\
 G_1(\chi)& = F_2(\chi)\, ,\\
 G_2(\chi)&= a_{22} \lambda(\chi)^2 \left(2+a_{22}\lambda(\chi)^2\right)\, ,
 \\
 H(\chi)&= 1-2\lambda(\chi)+a^2\lambda(\chi)^2\, , \\
 {\cal P}(\chi)&=\sqrt{A_5(\chi)J_e(\chi)}\, .
\end{align*}

\begin{widetext}

The function $J_e(\chi)$ is defined by:
\begin{equation*}
    J_e(\chi)= (1-E^2)(1-e^2)
    -2{\tilde \lambda(\chi)}
    \left[
      J_1
    +{\tilde \lambda(\chi)}^2J_3
    +{\tilde \lambda(\chi)}^4J_5
    +{\tilde \lambda(\chi)}^6J_7
    \right]+
    {\tilde\lambda(\chi)}^2\left[
    J_2
    +{\tilde \lambda(\chi)}^2J_4
    +{\tilde \lambda(\chi)}^4J_6
    +{\tilde \lambda(\chi)}^6J_8
    \right]
\end{equation*}
where the various functions $J_i$, with $i=1,2,..,8$ are:
\begin{align*}
J_1&=\frac{1-e^2}{p}-(1-E^2),\\ 
J_2&=\frac{a^2(1-E^2)+L_z^2+\tilde{Q}}{p^2}(1-e^2)-\frac{4}{p}+(1-E^2)\left(\frac{3+e^2}{1-e^2}\right)\,,\\
J_3&=\frac{[(L_z-a E)^2+\tilde{Q}-\epsilon_3/2+a_{13}E^2]}{p^3}(1-e^2)-\frac{a^2(1-E^2)+L_z^2+\tilde{Q}}{p^2}+\frac{1}{p}\left(\frac{3+e^2}{1-e^2}\right)-2(1-E^2)\frac{1+e^2}{(1-e^2)^2},\\
J_4&=\frac{a^2\tilde{Q}-2\epsilon_3+2a a_{22}EL_z}{p^4}(1-e^2)-\frac{4[(L_z-a E)^2+\tilde{Q}-\epsilon_3/2+a_{13}E^2]}{p^3}+\frac{a^2(1-E^2)+L_z^2+\tilde{Q}}{p^2}\left(\frac{3+e^2}{1-e^2}\right)\\
&-\frac{8}{p}\frac{1+e^2}{(1-e^2)^2}+(1-E^2)\frac{5+10e^2+e^4}{(1-e^2)^3},\\
J_5&=\frac{a a_{13}E(2 a E-L_z)-\epsilon_3a^2/2}{p^5}(1-e^2)-\frac{a^2\tilde{Q}-2\epsilon_3+2a a_{22}EL_z}{p^4}\\
&+\frac{[(L_z-a E)^2+\tilde{Q}-\epsilon_3/2+a_{13}E^2]}{p^3}\left(\frac{3+e^2}{1-e^2}\right)-2\frac{a^2(1-E^2)+L_z^2+\tilde{Q}}{p^2}\frac{1+e^2}{(1-e^2)^2}+\frac{1}{p}\frac{5+10e^2+e^4}{(1-e^2)^3}\\
&-(1-E^2)\frac{(1+3e^2)(3+e^2)}{(1-e^2)^4}\\
J_6 & =-\frac{a_{13}^2E^2+2a_{22}a^2L_z(L_z-a E)}{p^6}(1-e^2)-\frac{4 a a_{13}E(2 a E-L_z)-2 \epsilon_3a^2}{p^5}\\
&+\frac{a^2\tilde{Q}-2\epsilon_3+2a a_{22}EL_z}{p^4}
\left(\frac{3+e^2}{1-e^2}\right)-\frac{8 [(L_z-a E)^2+\tilde{Q}-\epsilon_3/2+a_{13}E^2]}{p^3}\frac{1+e^2}{(1-e^2)^2}\\
&+\frac{a^2(1-E^2)+L_z^2+\tilde{Q}}{p^2}\frac{5+10e^2+e^4}{(1-e^2)^3}-\frac{4}{p}\frac{(1+3e^2)(3+e^2)}{(1-e^2)^4}+(1-E^2)\frac{7+35e^2+21e^4+e^6}{(1-e^2)^5},\\
J_7 & = \frac{a_{13}a^3E (a E-L_z)-a a_{13}a_{22}E L_z }{p^7}(1-e^2)
+\frac{a_{13}^2E^2+2a_{22}a^2L_z(L_z-a E)}{p^6}\\
&+\frac{a a_{13}E(2 a E-L_z)-\epsilon_3a^2/2}{p^5}
\left(\frac{3+e^2}{1-e^2}\right)
-2\frac{a^2\tilde{Q}-2\epsilon_3+2a a_{22}EL_z}{p^4}\frac{1+e^2}{(1-e^2)^2}\\
&+\frac{[(L_z-a E)^2+\tilde{Q}-\epsilon_3/2+a_{13}E^2]}{p^3}\frac{5+10e^2+e^4}{(1-e^2)^3}
-\frac{a^2(1-E^2)+L_z^2+\tilde{Q}}{p^2}\frac{(1+3e^2)(3+e^2)}{(1-e^2)^4}\\
&+\frac{1}{p}\frac{7+35e^2+21e^4+e^6}{(1-e^2)^5}-4(1-E^2)\frac{(1+e^2)(1+6e^2+e^4)}{(1-e^2)^6},\\
J_8 & =-\frac{2 a^2 a_{13}^2E^2+a^2a_{22}^2L_z^2}{p^8}(1-e^2)-
4\frac{a_{13}a^3E (a E-L_z)-a a_{13}a_{22}E L_z }{p^7}\\
&-\frac{a_{13}^2E^2+2a_{22}a^2L_z(L_z-a E)}{p^6}\left(\frac{3+e^2}{1-e^2}\right)
-4\frac{2 a a_{13}E(2 a E-L_z)-\epsilon_3a^2}{p^5}\frac{1+e^2}{(1-e^2)^2}\\
&+\frac{a^2\tilde{Q}-2\epsilon_3+2a a_{22}EL_z}{p^4}\frac{5+10e^2+e^4}{(1-e^2)^3}
-\frac{4 [(L_z-a E)^2+\tilde{Q}-\epsilon_3/2+a_{13}E^2]}{p^3}\frac{(1+3e^2)(3+e^2)}{(1-e^2)^4}\\
&+\frac{a^2(1-E^2)+L_z^2+\tilde{Q}}{p^2}\frac{7+35e^2+21e^4+e^6}{(1-e^2)^5}
-\frac{16}{p}\frac{(1+e^2)(1+6e^2+e^4)}{(1-e^2)^6}+(1-E^2)\frac{9+84e^2+126e^4+36e^6+e^8}{(1-e^2)^7}.
\end{align*}

\end{widetext}

\end{appendix}

\bibliography{biblio}

\end{document}